\documentstyle[12pt]{article}
\newlength{\dinwidth}
\newlength{\dinmargin}
\newcommand{\ba}{\begin{array}}
\newcommand{\ea}{\end{array}}
\newcommand{\be}{\begin{equation}}
\newcommand{\ee}{\end{equation}}
\newcommand{\bea}{\begin{eqnarray}}
\newcommand{\eea}{\end{eqnarray}}

\def\bee{\begin{eqnarray}}
\def\eee{\end{eqnarray}}
\def\be{\begin{equation}}
\def\ee{\end{equation}}

\begin{document}
\thispagestyle{empty}
\addtocounter{page}{-1}
\begin{flushright}
IASSNS-HEP 97/29\\
SNUTP 97-042\\
{\tt hep-th/9705132}\\
\end{flushright}
\vspace*{1.3cm}
\centerline{\Large \bf M(atrix) Theory on ${\bf T}_5/{\bf Z}_2$ 
Orbifold and Five-Branes
\footnote{
Work supported in part by the NSF-KOSEF Bilateral Grant,
KOSEF Purpose-Oriented Research Grant 94-1400-04-01-3 and SRC-Program, 
Ministry of Education Grant BSRI 97-2418, the Monell Foundation
and the Seoam Foundation Fellowships, SNU Research Fund, and the Korea
Foundation for Advanced Studies Faculty Fellowship.}}
\vspace*{1.2cm} \centerline{\large\bf Nakwoo Kim${}^a$ and
Soo-Jong Rey${}^{a,b}$}
\vspace*{1.0cm}

\centerline{\large
\it Physics Department, Seoul National University,
Seoul 151-742 KOREA${}^a$}
\vspace*{0.3cm}
\centerline{\large \it School of Natural Sciences, Institute for
Advanced Study}
\centerline{\large\it Olden Lane, Princeton NJ 08540 USA${}^b$}
\vspace*{1.5cm}
\centerline{\Large\bf Abstract}
\vspace*{0.4cm}
We study M(atrix) theory description of M theory compactified on
${\bf T}_5/{\bf Z}_2$ orbifold. 
In the large volume limit we show that M theory dynamics is described by 
${\cal N} = 8$ supersymmetric USp(2N) M(atrix) quantum mechanics. 
Via zero-brane parton scattering, we show that each orbifold fixed point
carries anomalous G-flux $\oint [G/2 \pi]= - 1/2$. To cancel the anomalous
G-flux , we introduce twisted sector consisting of sixteen five-branes 
represented by fundamental representation hypermultiplets.
In the small volume limit we show that M theory dynamics is effectively
described by
by (5+1)-dimensional (8,0) supersymmetric USp(2N) chiral gauge theory. 
We point out that both perturbative and global gauge anomalies are 
cancelled by the sixteen fundamental representation hypermultiplets in 
the twisted sector. We show that M(atrix) theory is capable of turning on
spacetime background with the required sixteen five-branes out of zero-brane 
partons as bound-states. We determine six-dimensional spacetime spectrum from 
the M(atrix) 
theory for both untwisted and twisted sectors and find a complete agreement 
with the spectrum of (2,0) supergravity. 
We discuss M(atrix) theory description of compactification moduli space, 
symmetry enhancement thereof as well as further toroidal compactifications. 
\vspace*{1.1cm}

%\centerline{Submitted to Nuclear Physics B}

\newpage

%%%%%%%%%%%%%%%%%%%%%%%%%%%%%%%%%%%%%%%%%%%%%%%%%%%%%%%%%%%%%%%%%%%%%%%%%%%%%
\section{Introduction}
%%%%%%%%%%%%%%%%%%%%%%%%%%%%%%%%%%%%%%%%%%%%%%%%%%%%%%%%%%%%%%%%%%%%%%%%%%%%%
\setcounter{equation}{0}
Witten~\cite{witten95} has made an important observation that in the strong
coupling limit all known
superstring theories are most accurately described
by M-theory. The M-theory is a theory with manifest
eleven-dimensional Lorentz invariance and, when truncated to massless
excitations, reduces to the eleven-dimensional supergravity with 
appropriate gravitational corrections to the Chern-Simons coupling. 
While enormous progress has been
achieved based on analysis at the supergravity level, a satisfactory
microscopic description of the M-theory was missing.

Recently, Banks {\sl et.al.}~\cite{bfss} have put forward an 
interesting nonperturbative proposal of light-front M-theory,
so-called M(atrix) theory. By
extrapolating Type IIA string into strong coupling regime and boosting
infinitely along the `quantum' M-theory direction, 
they have identified constituent partons with zero-branes and Chan-Paton
gauge fields~\cite{wittenbound}.
Light-front dynamics of these partons is governed by
${\cal N} = 16$ supersymmetric M(atrix)
 quantum mechanics with SU(N) gauge group
and SO(9) R-symmetry~\cite{gaugeqm}.
Implicit to their proposal are two crucial assumptions: non-anomalous
$SO(10,1)$ Lorentz invariance and existence of a threshold
bound-state of the zero-brane partons. Despite these yet-unproven dynamical
assumptions and intrinsically light-front description, the M(atrix) theory
has successfully passed numerous consistency checks so far.
Furthermore, for finite $N$, the M(atrix) theory is shown to correspond
to discrete light-cone quantization (DLCQ) of M-theory~\cite{dlcq}. 

Among the most important issues in M(atrix) theory is a comprehensive
understanding of nontrivial compactifications. So far, toroidal
compactifications have been investigated mostly~\cite{
toroidal}. Being a perfectly
smooth manifold, however, the toroidal compactifications may not reveal
hidden intrinsically quantum and non-perturbative M(atrix) theory aspects.
In fact, it is not unlikely that
intrinsically M(atrix) theoretic aspects are tied with compactifications on
nontrivial spaces. Within the approach based strictly on low-energy
supergravity, the strongly coupled $E_8 \times E_8$ heterotic string
has already provided
such an example~\cite{horavawitten}: the strong coupling limit of
heterotic string is
described by the M-theory compactified on an orbifold ${\bf S}_1/{\bf Z}_2$.
Located at the two orbifold fixed points are nine-branes or `end of the
worlds'. The $E_8$ gauge multiplets arise as spectra
of the twisted sector localized on the nine-brane world-volume and are
the essential elements for satisfying the M-theory field equations.
If it were to offer a non-perturbative definition of the
M-theory, the M(atrix) theory should provide an equally
 consistent quantum mechanical
definition of ${\bf S}_1/{\bf Z}_2$ orbifold compactification and
produce the same result as the supergravity analysis at the least.
In the previous paper~\cite{kimrey}, we have investigated this issue and have found
M(atrix) theory definition of the orbifold compactification as well as the
correct twisted sector spectra.
Furthermore, based on the same M(atrix) theory compactification, various
dualities of the heterotic -- Type I strings have been understood in 
detail~\cite{kabatrey}.

In this paper, we extend our earlier work on M(atrix) theory on orbifold
and study ${\bf T}_5/{\bf Z}_2$ compactification in depth.
Witten~\cite{wittent5} and Dasgupta and Mukhi~\cite{dasguptamukhi} have
studied the same orbifold compactification of low-energy eleven-dimensional
supergravity and have shown
that the compactification vacua give rise to anomaly-free $(2,0)$ supergravity
spectra at low-energy. Despite the analysis was strictly based on the
eleven-dimensional supergravity on a singular orbifold, their results have
revealed a very interesting non-perturbative aspect of the M theory. 
Located at the $2^5$
orbifold fixed points are anomalous half-units of G-flux for
integral cohomology class $[G/2\pi] = -1/2$,
where $G_{\rm [MNPQ]} \equiv \nabla_{\rm [M} C_{\rm NPQ]}$
is the four-form tensor field strength of M theory.
In order to maintain vanishing total G-flux on ${\bf T}_5/{\bf Z}_2$,
sixteen five-branes has to be put on ${\bf R}_{5,1}$. This gives rise to
sixteen $(2,0)$ tensor multiplets as the twisted sector spectrum. The
total spectrum turns out  anomaly--free and in fact agrees with
the spectrum of type IIB string compactified on $K3$.

Given that the result of Witten and of Dasgupta and Mukhi has already 
revealed nonperturbative aspects of the M theory, a complete account of
${\bf T}_5/{\bf Z}_2$ orbifold compactification should therefore serve as 
a highly 
nontrivial test to the M(atrix) theory itself. Moreover, if the
M(atrix) theory is a complete descrition of M theory compactification
as well as dynamics on it,
fully consistent {\sl spacetime background} including nontrivial five-branes
on it as well as the complete {\sl spacetime} spectrum should emerge 
dynamically out of the M(atrix) theory.
As we will show, this indeed turns out to be the case. 
In particular, we show that 
the needed sixteen five-branes are dynamically arranged as
bound-states of zero-brane partons. This is precisely along the spirit of
the M(atrix) theory proposal: Spacetime background is a coherent 
configuration of M theory massless excitations. If the latter are to be
described by zero-brane bound-states, so should be the case for the 
spacetime backgrounds.

In Section 2, we begin with M(atrix)
theory description of M theory compactified on a large 
${\bf T}_5/{\bf Z}_2$ 
orbifold. By analyzing the action of orbifold projection to the 
Chan-Paton factors and R-symmetry, we show that the M(atrix) theory
relevant for the ${\bf T}_5/{\bf Z}_2$ compactification is given by
${\cal N} = 8$ supersymmetric M(atrix) quantum mechanics with gauge
group USp(2N) where N denotes the number of zero-brane partons. 
Via parton scattering off an orbifold fixed point, we probe presence
of anomalous G-flux of $- 1/2$ units emanating from each fixed points.
Consistency of the orbifold compactifiction requires cancellation of 
total $-16$ units of anomalous G-flux. This determines uniquely 
the twisted sector spectrum to consist of thirty-two `half' 
hypermultiplets in fundamental representation of USp(2N). 

In Section 3, we study M(atrix) theory description in the opposite 
limit, viz. M theory compacfied on a small ${\bf T}_5/{\bf Z}_2$ 
orbifold. The M(atrix) theory relevant in this limit may be derived
in two different ways. One can start from the ${\cal N}=8$ M(atrix)
quantum mechanics and T-dualize along all five orbifold directions. 
The method has been utilized previously for the case of a small
${\bf S}_1/{\bf Z}_2$ orbifold compactification, and exactly the same 
procedure applies in the present context. Alternatively, one may 
start with a covering space torus ${\bf T}_5$ in the small volume
limit. The limit is described by ${\cal N}=16$ supersymmetric Yang-Mills 
theory on a dual torus ${\tilde {\bf T}}_5$. One then apply the ${\bf 
Z}_2$ orbifold projection to obtain orbifold M(atrix) theory relevant 
for small volume limit. Since the first approach is a straightforward
exercise of the ${\bf S}_1/{\bf Z}_2$ orbifold compactification, we
explain the second approach in more detail. We show
that the prescription gives rise to (0,8) supersymmetric USp(2N) chiral
gauge theory on the dual torus ${\tilde {\bf T}}_5$ 
with one antisymmetric tensor representation hypermultiplet
plus twisted sector consisting of thirty-two `half' fundamental 
representation hypermultiplets. With these twisted sector spectrum, we
show that the M(atrix) gauge theory
is free from perturbative gauge and nonperturbative global anomalies.
Via the T-duality of the first 
approach, we then show that the the M(atrix) gauge theory is nothing
but the world-volume gauge theory of N overlapping small instantons of 
SO(32) Type I string theory. 
We also show that, degenerate ${\bf T}_5
/{\bf Z}_2$ orbifold limit in which one of the five radii shrinks to 
zero, the M(atrix) gauge theory is better interpreted as N overlapping
small instantons of $E_8 \times E_8$ heterotic string theory.     

One of the physically most appealing features of the M(tarix) theory is 
that all spacetime physical excitations are built out of bound-states 
of zero-brane partons. In particular, the sixteen longitudinal 
five-branes necessary for a consistent orbifold compactification can be
achieved via Landau-orbiting zero-brane parton configurations. In section
4, we show how the five-branes can be arranged and how the twisted sector
spectrum is induced by the five-brane background. We then deduce 
${\bf R}_{5,1}$ {\sl spacetime} massless excitations as well as BPS
spectra based entirely on the M(atrix) theory construction. We find a 
complete agreement with the (2,0) chiral supergravity multiplet 
structure and the BPS spectra symmetry SO(5,20; ${\bf Z}$).

In section 5, we study further toroidal compactification, viz. M theory
compactification on ${\bf T}_5/{\bf Z}_2 \otimes {\bf T}_m$. We show
that the small volume limit is described by a 
M(atrix) gauge theory on a dual orbifold ${\tilde {\bf T}}_5 
\otimes {\tilde {\bf T}}_m/{\bf Z}_2$ in which bulk theory is coupled to 
`impurity' boundary theory. In the bulk, the M(atrix) 
gauge theory is given by (5+m+1)-dimensional ${\cal N} = 16$ 
supersymmetric Yang-Mills theory of gauge group U(N). Near the 
thirty-two orbifold fixed points, the M-theory dynamics is described
by (5+1)-dimensional (0,8) supersymmetric USp(2N) chiral gauge 
theories. The  
thirty-two twisted sector `half' fundamental representation 
hypermultiplets are distributed among the thirty-two (5+1)-dimensional 
chiral gauge theories. Different distribution corresponds to T-dual
configuration of turning on Wilson lines either in heterotic $E_8 
\times E_8$ or Type I SO(32) string theories. Consistency of local
description requires cancellation of (5+1)-dimensional gauge anomalies
at each orbifold boundaries. We show that this is achieved via the 
Wess-Zumino term and anomaly inflows from the bulk thereof.

%%%%%%%%%%%%%%%%%%%%%%%%%%%%%%%%%%%%%%%%%%%%%%%%%%%%%%%%%%%%%%%%%%%%%
\section{M(atrix) Theory on a Large ${\bf T}_5/{\bf Z}_2$ Orbifold}
%%%%%%%%%%%%%%%%%%%%%%%%%%%%%%%%%%%%%%%%%%%%%%%%%%%%%%%%%%%%%%%%%%%%%%
\setcounter{equation}{0}
We begin with a large volume limit of the ${\bf T}_5 /{\bf Z}_2$ orbifold.
In this case, one should be able to describe the M theory parton dynamics 
accurate enough by  
a (0+1)-dimensional M(atrix) theory. 
However, the ${\bf Z}_2$ involution
is not acting freely on ${\bf T}_5$ but generates $2^5 = 32$ fixed 
points. As such, as in string theories, nontrivial effects of the 
orbifold compactification are expected to be present as one approaches 
any of these orbifold fixed points. 

In this section we show that, near the orbifold fixed point where 
the orbifold geometry is locally ${\bf R}_5/{\bf Z}_2$,
 M-theory parton dynamics is described by ${\cal N}=8$ supersymmetric 
USp(2N) M(atrix) quantum mechanics. Moreover,
via parton scattering off the orbifold fixed point, we probe the
presence of anomalous magnetic G-flux 
$[G/2\pi]$ of $- 1/2$ unit emanating from each orbifold fixed point. 
Since ${\bf T}_5/{\bf Z}_2$ is a compact space with
$2^5 = 32$ fixed points, extra background carrying +16 units of 
magnetic G-flux has to be turned on in order for the orbifold 
compactification to be compatible with G-flux conservatioG-flux conservation. 
The needed compensating fluxes are provided precisely by 
turning on sixteen longitudinal five-branes on the noncompact spacetime
${\bf R}_{5,1}$. These sixteen longitudinal five-branes in turn affects
the zero-brane parton dynamics in such a way manifestly consistent with
M(atrix) theory is defined. We show that the effects of sixteen longitudinal
five-branes are represented by twisted sector consisting of thirty-two 
`half' fundamental representation hypermultiplets of USp(2N). 

%%%%%%%%%%%%%%%%%%%%%%%%%%%%%%%%%%%%%%%%%%%%%%%%%%%%%%%%%%%%%%%%%%%%%%%%%%%%%
\subsection{M(atrix) Theory on ${\bf R}_5/{\bf Z}_2$ 
via Orbifold Projection}
%%%%%%%%%%%%%%%%%%%%%%%%%%%%%%%%%%%%%%%%%%%%%%%%%%%%%%%%%%%%%%%%%%%%%%%%%%%%%
We begin by recalling general definition of M(atrix) theory on a 
large orbifold~\cite{kimrey}.
In M(atrix) theory, orbifold compactification is defined in terms
of appropriate involution.
From the covering space point of view, the orbifold is defined by
starting with $N$ original and $N$ `image' 0-brane partons on 
the covering space, then subsequently projecting out with involution.
Therefore, in this description, coordinate and spinor matrices are
$2N \times 2N$ in size. Consider the fixed point of ${\bf R}_5/{\bf Z}_2$
located at 
$X_5 = X_6 = \cdots = X_9 = 0$. The Chan-Paton condition acting on 
M(atrix) theory coordinates is a discrete
gauge symmetry transformation, which is a direct product of orientation
reversal and parity transformation on ${\bf R}_5$. The condition is
given by~\cite{kimrey}:
\bee
{\bf X}_{||}  &=& \, + \, \,\, M \cdot
      {\bf X}^{\rm T}_{||} \cdot M^{-1}, \nonumber \\
{\bf X}_\perp  &=&  \,
- \,\, \, M \cdot {\bf X}^{\rm T}_\perp \cdot M^{-1}, \nonumber \\
{\bf \Theta} &=& \Gamma_\perp \, M \cdot {\bf \Theta}^{\rm T}
\cdot M^{-1}, \hskip1.5cm \Gamma_\perp = \Gamma_5 \cdots \Gamma_9.
\label{chanpatoncondition}
\eee
In M(atrix) theory, the two-branes are Landau-orbiting collective
excitations of the zero-brane partons. The two-brane charge is defined
by the first Chern class ${\cal Z}_2 :=
[F/2 \pi] = c_1 (G)$ of gauge group $G$.
As such, the above Chan-Paton condition inverts the two-brane orientation,
hence, sign of the charge for {\sl every} $x_\perp = (x_5, \cdots, x_9)$
directions. In terms of eleven-dimensional supergravity, the flip of
the first Chern class $[F/2 \pi]$ corresponds
to a simultaneous reversal $C_{\rm MNP} \rightarrow - C_{\rm 
MNP}$ for every
${}_{\rm MNP} = (5, \cdots , 9)$.

The matrix $M$ relates the original and the `image' 0-brane partons.
Hermiticity of the coordinate and the spinor matrices require $
M^{\rm T} \cdot M = \pm {\bf I}_{2N \times 2N}$. Possible choices of $M$
compatible with this condition can be written as $M_{2N \times 2N} =
{\bf I}_{N \times N} \otimes \sigma_\mu$.
In the previous work~\cite{kimrey}, we have shown that there are two 
possible choices with orbifold conditions and
embedding of the gauge symmetries. For $\mu = 0, 1, 3$, the compatible
gauge group of the M(atrix) theory turns out SO(2N) $\subset$ SU(2N),
while for $\mu = 2$, it was USp(2N) $\subset$ SU(2N).
In both choices of the M(atrix) theory gauge group, the components 
${\bf X}_\perp$'s transform as adjoint representation, viz.
two-index antisymmetric for SO(2N) and symmetric for USp(2N). 
We combine them with the M(atrix) theory gauge field $A_0$ and define
a gauge field:
\be
A_M = (A_0, \cdots, A_5) \equiv (A_0 , {\bf X}_5, {\bf X}_6, \cdots,
{\bf X}_9). 
\ee
The parallel components ${\bf X}_{||}$ transform as
two-index symmetric representations for
SO(2N) and anti-symmetric representations for USp(2N) respectively. 
We choose the M(atrix) theory gauge group to be USp(2N). In what
follows, through various consistency checks, we will find that 
this is indeed the correct choice of the M(atrix) gauge group. 

The spinor field $\bf \Theta$ decomposes into two eigenstates of 
$\Gamma_\perp$. Accordingly, the R-symmetry is broken to 
Spin(4)$\otimes$Spin(5) $\subset$ Spin(9). 
We adopt a representation of the sixteen-dimensional gamma matrices
so that the chirality is chosen
\be
\Gamma^{(11)} \equiv \Gamma^0 \Gamma^1 \Gamma^2 \cdots \Gamma^9 = 1.
\ee
In the decomposition of Spin(5)$\otimes$Spin(4)$ \subset$Spin(9)
we represent the 16$\times$16 gamma matrices as:
\bee
\Gamma^i &=& {\bf I}_{4 \times 4} \otimes (\gamma_i)_{4 \times 4}
 \hskip1.2cm (i = 1, 2, 3, 4)
\nonumber \\
\Gamma^{i + 4} &=& (\gamma_i)_{4 \times 4} \otimes (\gamma_5)_{4 \times 4}
 \hskip1cm (i=1,2,3,4,5)
\eee
where
\bee
\gamma_i &=& \left( \begin{array}{cc} 0 & + \sigma_i \\
- \sigma_i & 0 \end{array} \right), \hskip0.75cm (i = 1, 2, 3)
\nonumber \\
\gamma_4 &=& \left( \begin{array}{cc} {\bf I} & 0 \\ 0 & {\bf I} 
\end{array} \right),
\hskip1cm
\gamma_5 = 
\left( \begin{array}{cc} + i{\bf I} & 0 \\ 0 & - i{\bf I} \end{array} \right).
\eee
These choices are made for consistency with the chirality convention so that
\be
\gamma_1 \cdots \gamma_5 = 1, \hskip1cm \gamma_5^4 = 1.
\ee
Fermions are decomposed into two inequivalent multiplets. Decomposing
the R-symmetry Spin(4) = SU(2)$\times$SU(2), we denote ${\bf S}_a$ and
${\bf S}_{\dot a}$ as the right-handed and the left-handed spinors
and Dirac spinor of Spin(5) at the same time. Therefore,
the sixteen-component Majorana spinor can be decomposed into two
inequivalent chiral spinors of the R-symmetry:
\be
{\bf \Theta} = ({\bf 1}, {\bf 2}, {\bf 4}) \oplus
({\bf 2}, {\bf 1}, {\bf 4}) \equiv {\bf S}_a \oplus {\bf S}_{\dot a},
\ee
where
\bee
\Gamma_\perp {\bf S}_a &=& - {\bf S}_a \nonumber \\
\Gamma_\perp {\bf S}_{\dot a } &=& + {\bf S}_{\dot a}.
\eee
From the last condition in Eq.~(\ref{chanpatoncondition}), we 
then find that ${\bf S}_a$ transforms as an adjoint represetntion of
USp(2N) M(atrix) gauge group, while ${\bf S}_{\dot a}$ transforms
as an antisymmetric representation.

The fact that the two eight-component spinors transform differently
under the M(atrix) gauge group entails certain mismatch at one-loop.
Indeed, as we will show in the next sub-sections, the orbifold M(atrix)
quantum mechanics is consistent only after a twisted sector is included.
The twisted sector consists of thirty-two hypermultiplets transforming
as `half' fundamental representations under the gauge group USp(2N).

The field content of M(atrix) quantum mechanics is then given by:
\vskip0.5cm
\begin{tabular}{|c|c|c|c|c|c|} \hline
Sector & Multiplet & Bosons & Fermions & Spin(5)$\otimes$Spin(4) 
& USp(2N) \\ \hline
untwisted & gauge & $A_I$ & $ {\bf S}_a $ & $ ({\bf 5}, {\bf 1}) ; 
({\bf 4}, {\bf 2}_{\rm L})$
 & 2N(2N+1)/2 \\ \hline
 & antisymmetric & ${\bf X}^i$ & ${\bf S}_{\dot a}$ 
& $({\bf 1}, {\bf 4}) ; ({\bf 4}, {\bf 2}_{\rm R})$ &
2N(2N-1)/2
\\ \hline
twisted & fundamental & $ \phi^i_A $ & $\chi_{{\dot a} A}$ & 
$({\bf 1}, {\bf 4}); ({\bf 4}, {\bf 2}_{\rm R})$ & 2N
\\ \hline
\end{tabular}
\vskip0.5cm

The untwisted sector of the orbifold M(atrix) theory Lagrangian is given by
\bee
L_{\rm untwisted} = {\rm Tr}
\Big(\!\!\!\!&&\!\!\!\!
{1 \over 2 R} (D_t {\bf X}^i)^2 + {1 \over 2 R} (D_t A^I)^2
+ {\bf S}_a D_t {\bf S}_a + {\bf S}_{\dot a} D_t {\bf S}_{\dot a}
\nonumber \\
&& \nonumber \\ 
&+& {R \over 4} [{\bf X}^i, {\bf X}^j]^2
+ {R \over 4} [A^I, A^J]^2 + {R \over 2} [A^I, {\bf X}^j]^2
\nonumber \\
&& \nonumber \\
&+& 
2 i R {\bf X}^i \sigma^i_{a \dot a} \{ {\bf S}_a , {\bf S}_{\dot a} \}
- R {\bf S}_a \gamma_I [A^I, {\bf S}_a]
+ R {\bf S}_{\dot a} \gamma_I [A^I, {\bf S}_{\dot a}] \Big).
\eee
The $\gamma_I$ matrices act on Spin(5) R-symmetry indices, which
are suppressed in the expression.

Likewise, the twisted sector Lagrangian is given by
\bee
L_{\rm twisted}
=  \Big[ \!\!\!\! && \!\!\!\! 
{1 \over R} | (D_t + A_I) \phi^i_A|^2 
+ {\overline \chi}_A (D_t + R \, A_I \gamma^I) \chi_A
\nonumber \\
&+& {R \over 2} \, {\overline \phi}^i_A [{\bf X}^i, {\bf X}^j] \phi^j_A
- R \, |{\overline \phi}^i_A \, \phi^i_B|^2 
\nonumber \\
&-& R \, {\overline \phi}^i_A {\bf S}_a \sigma^i_{a \dot a} \chi_{{\dot a} A}
-  R \, {\overline \chi}_{{\dot a} A}
\sigma^i_{\dot a a} {\bf S}_a \phi^i_A \,\,\,\, \Big].
\eee
The $A,B = 1, \cdots, 16$ indices denote the sixteen fundamental 
representation hypermultiplets obtained by adjoining pairs of
`half' fundamental representation hypermultiplets. 

The corresponding Hamiltonian is straightforwardly derived:
\bee
H_{\rm total} = R \Big[ {\rm Tr} \Big(
&+& {1 \over 2} {\bf \Pi}_i^2 + {1 \over 2} E_I^2 
\nonumber \\
&-&{1 \over 4} [{\bf X}^i, {\bf X}^j]^2 - {1 \over 4} [A_I, A_J]^2
- {1 \over 2} [A_I, {\bf X}^i]^2 
\nonumber \\
&-& 2 i {\bf X}^i \sigma^i_{a \dot a} \{ {\bf S}_a, {\bf S}_{\dot a} \}
+ {\bf S}_a \gamma_I [A^I, {\bf S}_a]
- {\bf S}_{\dot a} \gamma_I [A^I, {\bf S}_{\dot a} ] \Big)
\nonumber \\
&+& |\Pi_{iA}|^2 - {\overline \chi}_A A_I \gamma^I \chi_A
\nonumber \\
&-&{1 \over 2} {\overline \phi}^i_A \, [{\bf X}^i, {\bf X}^j] \, \phi^j_A
+ |{\overline \phi}^i_A \, \phi^i_B|^2
\nonumber \\
&+& {\overline \phi}^i_A {\bf S}_a \sigma^i_{a \dot a} \chi_{\dot a A}
+ {\overline \chi}_{\dot a A} \sigma^i_{\dot a a} {\bf S}_a 
\phi^i_A \,\,\,\,\, \Big].
\label{hamiltonian}
\eee
The overall dependence on the 11-th directional radius $R$ represent
correctly that Eq.(\ref{hamiltonian}) should be
interpreted as the light-cone Hamiltonian.

%%%%%%%%%%%%%%%%%%%%%%%%%%%%%%%%%%%%%%%%%%%%%%%%%%%%%%%%%%%%%%%%%%%%%%%
\subsection{Probing Anomalous G-Flux via Parton Scattering}
%%%%%%%%%%%%%%%%%%%%%%%%%%%%%%%%%%%%%%%%%%%%%%%%%%%%%%%%%%%%%%%%%%%%%%
The ${\bf T}_5/{\bf Z}_2$ orbifold has 
$2^5 = 32$ fixed points. It is likely that these fixed points are more 
than mere geometrical singular loci. Indeed, for 
${\bf S}_1/{\bf Z}_2$ orbifold~\cite{kimrey}, we have found that the
each fixed point carries $2^3 = 8$ units of D8--brane charge and 
create nonzero vacuum energy tadpole. In order to cancel the RR charge
and the vacuum energy, it was necessary to introduce eight complex
fermions in the fundamental representation of the gauge group. 

We now examine similar possibility for the ${\bf T}_5/{\bf Z}_2$ orbifold
fixed points using the zero-brane partons as probes.
Consider a neighborhood of one of the 32 fixed points. Locally 
the geometry is ${\bf R}_5/{\bf Z}_2$, 
an orbifold of noncompact five-dimensional space.
Place a zero-brane parton as a local probe.
A zero-brane parton scattering slowly off the fixed point will experience 
 presence of a mirror zero-brane parton moving opposite to
the probing zero-brane parton. As such we expect that the net effect of
orbifold fixed point is equivalent to a potential generated via forward
scattering between the two zero-branes.
Dynamics of a zero-brane moving toward the fixed point is 
described by USp(2) M(atrix) quantum mechanics. 
Consider the zero-brane parton moving in (8,9) plane with velocity $v$
and impact parameter $b$ relative to the orbifold fixed point. This 
is described by the background field configuration
$X_8 = vt \, \sigma_3/2 $ and $X_9 = b \, \sigma_3/2$.
Expanding around the background, it is straightforward to see that
bosonic and spinor fields in the rank-two antisymmetric representation 
are 
flat directions, reflecting translational invariance on ${\bf R}_{5,1}$.
Bosonic and spinor fields in the adjoint (symmetric)
representation give rise to massive fluctuations. In the covariant 
background field gauge, there are two bosonic modes with frequency
$\gamma^2 \tau^2 + b^2$, two bosonic modes with frequency $
\gamma^2 \tau^2 + b^2 \pm 2 \gamma$. Similarly, there are 
eight massive fermionic modes. Thus, the one-loop amplitude is given
by 
\bee
[{\cal A}_{\rm D0 - O4}]_{\rm USp} (\gamma \tau, b) 
&=& {\rm det}^{-2} (-\partial_\tau^2 + \gamma^2 \tau^2 + b^2)
\cdot {\rm det}^{-1} (-\partial_\tau^2 + \gamma^2 \tau^2 + b^2 + 2 \gamma)
\nonumber \\
&\cdot &
{\rm det}^{-1} (-\partial_\tau^2 + \gamma^2 \tau^2 + b^2 - 2 \gamma)
\cdot {\rm det}^{+4} \left( \begin{array}{cc}
\partial_\tau & \gamma \tau - i b \\
\gamma \tau + i b & \partial_\tau \\
\end{array} \right)
\label{oneloop}
\eee
Again, straightforward calculation yields the result as
\be
[{\cal A}_{\rm D0 - O4}]_{\rm USp} (r, v)
= v \int_0^\infty {d s \over \sqrt {\pi s}} 
\, e^{- sr^2} {1 + \cos 2 v s - 2 \cos v s \over \sin v s},
\ee
where $r^2 = b^2 + v^2 \tau^2$. Expanding the integrand in the small
velocity limit, we get the potential energy for a slow zero-brane parton
scattering off the orbifold fixed point (4-orientifold):
\be
[{\cal V}_{\rm D0 - O4}]_{\rm USp} (r, v) \, 
= \, -\, {v^2 \over 2 r^3} + {25 \over 32} {v^4 \over r^7} 
+ \cdots.
\label{orbifoldscattering1}
\ee

Suppose the M(atrix) gauge group were SO(2N) instead of USp(2N). In 
this case, the zero-brane parton scattering off the fixed point is 
described by untwisted sector one-loop amplitude of SO(2) M(atrix) 
quantum mechanics. In this case, massive modes are provided by the
symmetric representations. There are four bosonic massive modes
with frequency $\gamma^2 \tau^2 + b^2$ and eight fermionic massive
modes. The one-loop amplitude in this case is given by 
\be
[{\cal A}_{\rm D0 - O4}]_{\rm SO} (\gamma \tau, b)
= {\rm det}^{-4} (- \partial_\tau^2 + \gamma^2 \tau^2 + b^2)
\cdot {\rm det}^{+4} \left( \begin{array}{cc}
\partial_\tau & \gamma \tau - i b \\
\gamma \tau + i b & \partial_\tau \end{array} \right)
\ee 
Straightforward regularized calculation yields 
\be
[{\cal A}_{\rm D0 - O4}]_{\rm SO} (r, v) =
v \int_0^{+\infty} { d s \over \sqrt{\pi s}}
e^{- s r^2} {2 - 2 \cos v s \over \sin v s}.
\ee
We thus find the effective potential
\be
[{\cal V}_{\rm D0-O4}]_{\rm SO} ( r, v)
= + { v^2 \over 2 r^3} + { 5 \over 32} {v^4 \over r^7} + \cdots.
\label{orbifoldscattering2}
\ee

Let us contrast the differences between SO and USp choices of the
M(atrix) gauge group. For both cases, massive modes that contribute
to the one-loop amplitude, hence, to the effective potential arise
from the two-index symmetric representation. For G=USp(2N)
this is the adjoint representation supermultiplet and is interpreted
as the zero-brane parton coordinates on ${\bf T}_5/{\bf Z}_2$. 
For G=SO(2N) this is the symmetric representation hypermultiplet and 
is interpreted as the transverse coordinates of zero-brane partons
on noncompact spacetime ${\bf R}_{5,1}$. Since we have arranged the 
scattering to take place entirely on ${\bf T}_5/{\bf Z}_2$ but at 
a fixed position on ${\bf R}_{5,1}$, by translational invariance, 
the nontrivial one-loop effect should arise entirely from the 
fluctuations of ${\bf T}_5/{\bf Z}_2$ part coordinates of the zero-brane. 
This is precisely the case if we have chosen the M(atrix) gauge group
as USp(2N) but not for SO(2N). This argument provides one of the 
physical basis for our choice of USp(2N) as the correct M(atrix) theory
gauge group.

We now consider parton scattering off a longitudinal five-brane background.
The minimally charged five-brane is represented by a hypermultiplet
transforming as a fundamental representation under the M(atrix) theory 
gauge group~\footnote{Likewise, multiply charged longitudinal five-branes are 
represented by hypermultiplets in higher-dimensional representations
of the M(atrix) gauge group. In M(atrix) theory, these hypermultiplets
in arbitrary representation can be constructed by turning on appropriate
backgrounds of multiple longitudinal five-branes on ${\bf R}_{5,1}$ 
spacetime.  We will show the construction explicitly in section 4.2.} .
For both SO(2N) and USp(2N) gauge group choices, a simple counting of
the massive bosonic and fermionic modes in the background of 
$X_8 = v t \sigma_3/2$ and $X_9 = b \sigma_3 / 2$ shows that the 
one-loop scattering amplitude is given by the same expression.
For a unit-charged longitudinal five-brane background, it is 
\be
{\cal A}_{\rm D0 - L5} (\gamma \tau, b)
= {\rm det}^{-4} (-\partial_\tau^2 + {\gamma^2 \tau^2 \over 4}
+ {b^2 \over 4}) \cdot
{\rm det}^{+4} \left( \begin{array}{cc}
\partial_\tau & {\gamma \tau \over 2} - i {b \over 2} \\
{\gamma \tau \over 2} + i {b \over 2} & \partial_\tau \end{array}
\right)
\label{fivebranescatt}
\ee
This yields the potential energy
\be
{\cal V}_{\rm D0 - L5} (r, v)
= + {v^2 \over r^3} + {5 \over 4} {v^4 \over r^7} + \cdots.
\label{fivebranescattering}
\ee

Comparing the potential energy exerted by an orbifold fixed point
(four-orientifold) Eq.~(\ref{orbifoldscattering1})
to that exerted by a unit-charged longitudinal five-brane
Eq.~(\ref{fivebranescattering}),
we find that the orbifold fixed point carries positive one-half
or negative one-half units of the five-brane tension depending on whether 
M(atrix) gauge group is SO(2N) or USp(2N).
Since the brane charges are measured in units of tension, we interpret
this result as implying that the orbifold fixed point of SO(2N) respectively USp(2N) M(atrix) theory carries $\pm$ 1/2 units of five-brane charge. 
The $-1/2$ unit of anomalous G-flux emanating from the orbifold fixed point
was fist pointed out by Witten~\cite{wittent5} and, independently,
by Dasgupta and Mukhi~\cite{dasguptamukhi} based on spacetime gravitational
anomaly cancellation consideration. It is gratifying that the M(atrix) 
quantum mechanics can probe the anomalous G-flux correctly and consistently
with their results. In fact, the M(atrix) theory provides another means
of probing the anomalous G-flux via gauge anomaly cancellation, hence,
confirming results obtained by Witten~\cite{wittent5} and Dasgupta and 
Mukhi~\cite{dasguptamukhi}.
In the next section, we will also find that the USp(2N) is the correct
choice of the M(atrix) gauge group from perturbative and global 
gauge anomaly cancellation considerations in the shrinking 
${\bf T}_5/{\bf Z}_2$ limit.

Recalling that there are $2^5 = 32$ orbifold fixed points on 
${\bf T}_5/{\bf Z}_2$, we find that, only for the USp(2N) M(atrix) gauge
group choice but not for SO(2N), 
the total anomalous G-flux of $ 32 \times (-1/2) = - 16$ units can be 
cancelled by turning on sixteen longitudinal five-branes. Locally the
cancellation is incomplete unless the ${\bf T}_5/{\bf Z}_2$ orbifold 
is taken to degenerate limit in the compactification moduli space. 
For now, we will tacitly assume such a limit should do consistency of the 
compactification arise. We will elaborate more on this issue in section 5.

Note that, in deducing the anomalous G-flux of -1/2 unit from 
each orbifold fixed point, hence, total thirty-two multiplicity
of twisted sector hypermultiplets, 
seemingly trivial 1/2 factor difference of U(1) charge between
the fundamental and the two-index representations has played a crucial
role.  See Eqs.({\ref{oneloop}, 
\ref{fivebranescatt}). While this is a direct consequence of a simple group 
theoretic fact, it is instructive to understand the 1/2 factor 
difference from geometrical point of view as well. 
Let us recall the procedure
for calculating zero-brane parton scattering off the orbifold fixed
point and off a static object represented by the hypermultiplets. 
In the first case, using the method of image, we have arranged a
mirror zero-brane parton at the ${\bf Z}_2$ transformed position.
The net potential is then calculated in terms of relative distance
and relative velocity between the original zero-brane parton and the
mirror zero-brane. Note that this procedure yields {\sl twice} the 
distance between the original zero-brane parton and the orbifold
fixed position. At the same time the relevant mass that enters the
relative dynamics is the reduced mass, viz. {\sl half} of the
original zero-brane parton mass. 
In the case of zero-brane parton
scattering off the fundamental representation hypermultiplet, we have
arranged the hypermultiplet as a fixed background. Thus, the distance
as well as the reduced mass that enters the relative dynamics are 
geometrically measured ones. 
 
Therefore, if we denote the relative distance, velocity and potential
as $R, V, U$ between the probe zero-brane parton and the image parton
and as $r, v, {\cal V}$ between the probe zero-brane parton and the
orbifold fixed point, then the equation of motion for the probing
zero brane reads
\be
{1 \over 2} \ddot R = - {d U (R, V) \over d R}
\ee
in the first method of description and
\be
\ddot r = - {d {\cal V} (r, v) \over d r}
\ee
in the second case.
Since the two descriptions should be the same, using the fact that 
$R = 2 r, v = \dot x = V/2$, we find that
\be
{\cal V} (r, v)  = { 1 \over 2} U(R = 2r, V = 2 v).
\ee
This explains the group theory factor 1/2 difference in 
Eqs.(\ref{oneloop}, \ref{fivebranescatt}) more intuitively via 
elementary geometric consideration.
In the above results, we have taken the normalization that $r, v$ denote
the relative distance and velocity for the fundamental representation, viz.
a single longitudinal five-brane, and $2r, 2v$ for the two-index 
adjoint or antisymmetric representations.

%%%%%%%%%%%%%%%%%%%%%%%%%%%%%%%%%%%%%%%%%%%%%%%%%%%%%%%%%%%%%%%%%%%%%%%%%%%
\section{M(atrix) Gauge Theory on Small ${\bf T}_5/{\bf Z}_2$   }
%%%%%%%%%%%%%%%%%%%%%%%%%%%%%%%%%%%%%%%%%%%%%%%%%%%%%%%%%%%%%%%%%%%%%%%%%%%%
In this section, we study M theory compactification for small ${\bf T}_5/{\bf
Z}_2$ orbifold. The first method is to make Fourier transform the
M(atrix) quantum mechanics we have identified above. This is essentially
to take into account of all possible open string configurations connecting
among the zero-brane partons and their images around the compactified
directions. Since this is a straightforward extension of the method 
utilized already for ${\bf S}_1/{\bf Z}_2$ compacification~\cite{kimrey},
in this section, we explain alternative prescription for deriving the
same result. 

%%%%%%%%%%%%%%%%%%%%%%%%%%%%%%%%%%%%%%%%%%%%%%%%%%%%%%%%%%%%%%%%%%%%%%%%%
\subsection{M(atrix) Gauge Theory via Orbifold Projection}
%%%%%%%%%%%%%%%%%%%%%%%%%%%%%%%%%%%%%%%%%%%%%%%%%%%%%%%%%%%%%%%%%%%%%%%%%%%

We begin with M theory compactification on a small ${\bf T}_5$. In the
M(atrix) theory approach to M theory, this 
limit is most appropriately described by ${\cal N} = 16$ supersymmetric 
U(N) Yang-Mills theory on a dual torus ${\tilde {\bf T}}_5$.
M theory dynamics is then extracted from scattering of excitations in the 
moduli space of the gauge theory. 
Next consider M theory compactification on a small ${\bf T}_5/{\bf Z}_2$.
The ${\bf Z}_2$ involution is a combined
operation of parity transformation to the ${\bf T}_5$ coordinates and 
orientation reversal $C_{\rm MNP} \rightarrow - C_{\rm MNP}$. 
Accordingly, the M(atrix) gauge theory is obtained via ${\bf Z}_2$
involution of the ${\cal N}= 16$ supersymmetric U(N) Yang-Mills theory.

We begin with arranging mirror-image partons on ${\bf T}_5$. The total 
number of zero-brane partons is equal to 2N. Their dynamics
is described by ${\cal N}=16$ U(2N) Yang-Mills theory on a dual torus
${\tilde {\bf T}}_5$. In the vanishing volume limit, the field content
consists of a gauge multiplet $(A_\mu, {\bf S}_{\dot a})$ and 
a matter multiplet $({\bf X}^i, {\bf S}_a)$. Acting on these fields, 
the ${\bf Z}_2$ involution is a combination
of `parity' transformation $P$ and orientation reversal $\Omega$. 
The `parity' transformation acts as:
\bee
{\rm P} \hskip0.5cm : \hskip0.5cm 
A_\mu (x) &\rightarrow&   P \cdot A_\mu ( x) \cdot P^{-1}
\equiv \,\, + A_\mu (P \cdot x) \nonumber \\
{\bf X}^i (x) &\rightarrow& P \cdot {\bf X}^i ( x) \cdot P^{-1}
\equiv \,\, - {\bf X}^i (P \cdot x) \nonumber \\
{\bf \Theta}_\alpha (x) &\rightarrow& P \cdot {\bf \Theta}_\alpha 
(x) \cdot P^{-1} \equiv
\Gamma_\perp \, {\bf \Theta} (P \cdot x) \,\,\, .
\eee
The orientation reversal transformation $\Omega$ acts as:
\bee
\Omega \hskip0.5cm : \hskip0.5cm
A_\mu (x) &\rightarrow& \Omega \cdot A_\mu (x) \cdot \Omega^{-1}
\equiv \,\, \pm A^{\rm T}_\mu (x) 
\nonumber \\
{\bf X}^i (x) &\rightarrow& \Omega \cdot {\bf X}^i (x) \cdot \Omega^{-1}
\equiv \,\, \pm {\bf X}^{i \rm T} (x)
\nonumber \\
{\bf \Theta}_\alpha (x) &\rightarrow& \Omega \cdot 
{\bf \Theta}_\alpha^{\rm T} (x)
\cdot \Omega^{-1} \equiv  \pm \epsilon \, {\bf \Theta}^{\rm T}(x), 
\eee 
in which we have explicitly shown two-fold overall sign ambiguity and
two-fold $\epsilon = \pm$ relative sign ambiguity in the definition of
the orientation reversal transformation. In the present case, the latter
ambiguity is resolved by an appropriate choice of the six-dimensional 
spinor chirality convention.
The combined operations $\Pi \equiv \Omega \cdot P$ is a symmetry
of the U(2N) M(atrix) gauge theory, where the 
fields transform as:
\bee
A_\mu (x) &\rightarrow& \pm A_\mu^{\rm T} (P \cdot x)
\nonumber \\
{\bf X}^i (x) &\rightarrow& \mp \,\,\, {\bf X}^{i \rm T} (P \cdot x)
\nonumber \\
{\bf \Theta} (x) &\rightarrow& \pm \,\, \Gamma_\perp
\, {\bf \Theta}^{\rm T} (P \cdot x) .
\label{cpcondition}
\eee
The combined operation yields the ${\bf T}_5/{\bf Z}_2$ orbifold 
M(atrix) gauge theory so long as $P \cdot x = x$ itself. 
As in the M(atrix) quantum mechanics, the involution
acts both on the dual parameter space ${\tilde {\bf T}}_5$ and the gauge
group U(2N).  

The two-fold overall sign ambiguity in Eq.~(\ref{cpcondition}) corresponds
to two distinct choices of gauge group projected out of SU(2N). 
For the upper sign choice, Eq.~(\ref{cpcondition}) projects out the 
gauge fields into two-index symmetric representation, the pseudoscalar
fields into two-index antisymmetric representation and the sixteen 
component spinors into a sum of symmetric representation, eight component
spinor ${\bf S}_a$ and antisymmetric representation, eight component spinor
${\bf S}_{\dot a}$. Note that
two spinors have opposite six-dimensional chirality. This is precisely
the field content of a (5+1)-dimensional (0,8) supersymmetric USp(2N) chiral
gauge theory consisting of 
a gauge supermultiplet $(A_I, {\bf S}_a)$
and one hypermultiplet $({\bf X}^i, {\bf S}_{\dot a})$ 
in antisymmetric representation. For the lower sign choice, 
Eq.(\ref{cpcondition}) yields SO(2N) gauge group instead of USp(2N).
As we will see momentarily, this choice leads to an inconsistent theory
and we only concentrate on the USp(2N) gauge group choice from now on.

The above field content as it stands is anomalous. This is because
the involution projects out (5+1)-dimensional fermion content in 
a chirally asymmetric manner: one adjoint fermion of positive chirality
and one antisymmetric fermion of negative chirality.
The (5+1)-dimensional perturbative gauge anomaly is given by quartic 
Casimir operator $C_4 ({\tt R} ) \equiv {\rm Tr}_{\tt R} F^4$
of USp(2N) gauge group. With the above field content we find that the
anomaly measured in unit of left-handed fundamental representation
fermion is given by
\bee
X^8_{\rm untwisted}
 &=&  {\rm Tr}_{\tt adjoint} F^4  -  {\rm Tr}_{\tt antisymm}  F^4
\nonumber \\
&=&  \Big[ \{ (2N +8) {\rm tr} F^4 + 3 ({\rm tr} F^2)^2 \}
 - \{ (2N -8){\rm tr} F^4 + 3 ({\rm tr} F^2)^2 \}  \Big]
\nonumber \\
&=& 
16 \, {\rm tr} F^4.
\eee
The gauge anomaly is purely quartic. As such, unless the anomaly is cancelled
by adding appropriate chiral matter, the M(atrix) gauge theory is inconsistent.
By inspection, it is clear that by adding sixteen fundamental or, equivalently,
thirty-two `half' fundamental representation hypermultiplets the 
nonfactorizable anomaly can be cancelled. 
Had the M(atrix) theory gauge group been SO(2N) instead of USp(2N), the
symmetric and the antisymmetric representations switch their role as gauge
and adjoint representations. In turn, the resulting anomaly is minus to that
of the USp(2N) case. This anomaly cannot be cured neither by adding chiral
matters nor by Green-Schwarz mechanism. The theory makes sense only if 
it is coupled to (0,1) supergravity in six 
dimension~\cite{schwarz}~\footnote{Closely
related analysis but without consideration of potential 
global gauge anomalies has been made
independently by \cite{ferretti, smith}.} .

In fact, the USp(2N) gauge group
with one antisymmetric representation and thirty-two `half' fundamental
representation hypermultiplets is very special. This is the only (0,8)
gauge theory content in six dimensions which is free from perturbative
gauge anomalies for {\sl all} N. Since the rank of gauge group N is 
interpreted as the total number of zero-brane partons, hence, the 
size of longitudinal momentum in units of $1/R$, we find that the
anomaly consideration selects uniquely the USp(2N) M(atrix) gauge group
as a consistent choice. We also note that, when adjoined with SO(32) gauge 
group of the twisted sector, the M(atrix) gauge theory remains anomaly free
even after it is coupled to (0,1) supergravity in six 
dimensions~\cite{schwarz}. 

One needs to check further consistency condition.
For a single parton, N=1, the M(atrix) gauge theory group is 
USp(2) = SU(2), and there arises potential inconsistency due to 
nonperturbative global gauge 
anomaly. This is because $\Pi_6$(SU(2)) = ${\bf Z}_{12}$.
The consistency condition for absence of global gauge anomaly has been
performed recently. For USp(2) = SU(2), the number of fundamental
representation hypermultiplets has to be 4 modulo 6~\cite{bershadskyvafa}. 
The sixteen 
fundamental, hence, thirty-two `half' fundamental representations 
fits perfectly to the consistency condition. Putting together both
the perturbative and global gauge anomaly consistency conditions, we
thus conclude that the M(atrix) gauge theory with gauge group USp(2N) 
describes a consistent M-theory compactification on ${\bf T}_5/{\bf Z}_2$ 
only when a twisted
sector spectrum of thirty-two `half' fundamental representation
hypermultiplets are introduced. These hypermultiplets also exhibit
SO(32) global symmetry. Later they will be identified with the gauge
group of sixteen longitudinal five-branes located at the orbifold
fixed points.

Summarizing what we have found so far, field content and quantum numbers 
of the M(atrix) gauge theory are given by :
\vskip0.5cm
\begin{tabular}{|c|c|c|c|c|c|} \hline
Sector & Multiplet & (Components) & Spin(5)$\otimes$Spin(4) & USp(2N) 
& SO(32)  \\ \hline
untwisted & gauge & ($A_I; {\bf S}_{ a}$) & 
$({\bf 5}, {\bf 1} ; {\bf 4}, {\bf 2}_{\rm L})$  & 2N (2N + 1)/2 & 
${\bf 1}$
 \\ \hline
 & symmetric & $({\bf X}^i$; ${\bf S}_{\dot a})$ & ({\bf 1}, {\bf 4}; 
{\bf 4}, ${\bf 2}_{\rm R}$) &  2N(2N - 1)/2 & 
{\bf 1}
\\ \hline
twisted & fundamental & ( $\phi^i_A ; \chi_{\dot a A})$
 & $({\bf 1}, {\bf 4}; {\bf 4}, {\bf 2}_{\rm R}) $ & 2N & {\bf 32}
\\ \hline
\end{tabular}
\vskip0.5cm

The untwisted sector Lagrangian on ${\tilde {\bf T}}_5$ is given by
\bee
L_{\rm untwisted} = {1 \over g^2_{\rm YM}}
{\rm Tr} \int d^5 \sigma \, \Big[
&-& {1 \over 4} F_{\alpha \beta}^2 + (D_\alpha {\bf X}^i)^2
+ {1 \over 4} [{\bf X}^i, {\bf X}^j]^2 
\nonumber \\
&& \nonumber \\
&+& {\bf S}_a {\cal D}_{\rm R} {\bf S}_a 
+ {\bf S}_{\dot a} {\cal D}_{\rm L} {\bf S}_{\dot a}
+ 2 i {\bf X}^i \sigma^i_{a \dot a}
\{ {\bf S}_a, {\bf S}_{\dot a} \} \,\,\,\, \Big], 
\eee
where
the chiral covariant derivatives are
${\cal D}_{\rm L} = D_t - \gamma_I D_I $ and ${\cal D}_{\rm R} 
\equiv D_t + \gamma_I D_I$ respectively.

Similarly, the twisted sector Lagrangian is given by
\bee
L_{\rm twisted} 
= \int d^5 \sigma \, 
\Big[ &+& | D_\alpha \phi^i_A|^2 + 
{\overline \chi}_A {\cal D}_{\rm R} \chi_A
\nonumber \\
&& \nonumber \\
&-& {1 \over g^2_{\rm YM}} |{\overline \phi}^i_A \phi^i_B|^2 
- {\overline \phi}^i_A [{\bf X}^i, {\bf X}^j] \chi^j_A
\nonumber \\
&& \nonumber \\
&-& {\overline \phi}^i_A \sigma^i_{a \dot a} {\bf S}_a \gamma^i 
\chi_{{\dot a} A}
- \phi^i_A \sigma^i_{\dot a a } {\overline \chi}_{\dot a A} \gamma^i 
{\bf S}_a \,\,\,\, \Big].
\eee
\vskip0.3cm
%%%%%%%%%%%%%%%%%%%%%%%%%%%%%%%%%%%%%%%%%%%%%%%%%%%%%%%%%%%%%%%%%%%%%%%%%%%
\subsection{Probing Anomalous G-Flux via T-Duality}
%%%%%%%%%%%%%%%%%%%%%%%%%%%%%%%%%%%%%%%%%%%%%%%%%%%%%%%%%%%%%%%%%%%%%%%%%%%
It is possible to understand the M(atrix) gauge theory constructed above
in an intuitive way. In this sub-section, we show that the M(atrix) gauge 
theory is in fact the world-volume gauge theory of N overlapping SO(32) 
small instantons~\cite{wittensmallinstanton} of Type I string theory. 
This implies that dynamics of M-theory 
compactified on ${\bf T}_5/{\bf Z}_2$ orbifold can be understood via
scattering experiments of gauge theory excitations in the moduli space of 
N overlapping SO(32) small instanton world-volume gauge theory in the large 
N limit.
Moreover, this also enables to identify the origin of the anomalous G-flux
as T-dual of the nine-brane charge carried by orientifold in Type I string
theory. 

To proceed further, we take the 11-th direction as the `quantum' direction. 
The `quantum' direction is compactified on a circle of radius $R_{11}$ along 
which the M-theory is boosted. The radius $R_{11}$ will be taken to 
infinity in the end. Viewed this way, M(atrix) theory compactified on
transverse ${\bf T}_5/{\bf Z}_2$ is identified with Type IIA string theory 
compactified on ${\bf T}_5$ with 4-orientifolds (O4), which we will refer
as Type $\tilde{I}{}^\prime$ theory. 

On the compactified
${\bf T}_5/{\bf Z}_2$ are placed N D0-brane partons. The sixteen longitudinal 
five-branes become thirty-two D4-branes with ${\bf Z}_2$ symmetric 
identification. Label the compactified directions 
as $X_5, \cdots, X_9$ and their radii as $R_5, \cdot, R_9$. Then the 
4-orientifolds are located at $X_5 = 0, \pi R_5, 
\cdots,  X_9 = 0, \pi R_9$, 
viz. at the thirty-two fixed points of ${\bf T}_5/{\bf Z}_2$. 
Let us T-dualize all the compact directions. 
This maps the Type $\tilde{I}{}^\prime$
theory into Type I theory on a dual torus ${\tilde {\bf T}}_5$. 
The thirty-two D4-branes in the twisted sector turns into thirty-two 
D9-branes producing SO(32) Type I string gauge group. Likewise, the N 
D0-brane partons turns into N parallel D5-branes wrapped around the dual torus.

As is well-known, due to the absence of Coulomb branch, these D5-branes are 
SO(32) small instantons. Each individual small instantons carry USp(2)
= SU(2) world-volume gauge theory. When N small instantons overlap, the
gauge symmetry is enhanced to USp(2N). This is precisely the M(atrix)
gauge group we have identified earlier. Furthermore, the collective 
coordinates of small instantons along $X_1, \cdots, X_4$ directions are
identified with the non-compact transverse directions of infinitely
boosted M-theory, viz, M(atrix) theory. Due to Chan-Paton condition, it
is known that the collective coordinates in a two-index, antisymmetric
representation under USp(2N)~\cite{wittensmallinstanton}. 
They are precisely the matter multiplet
$({\bf X}^i, {\bf S}_a)$ of the M(atrix) gauge theory in the untwisted
sector. 

Finally, at the overlap of N D5-branes and thirty-two D9-branes,
there appear
thirty-two, `half' fundamental representation matter multiplets of 
USp(2N). Again they are precisely the twisted sector multiplets of the
M(atrix) gauge theory introduced in the previous section to ensure USp(2N) 
gauge anomaly cancellation.

%%%%%%%%%%%%%%%%%%%%%%%%%%%%%%%%%%%%%%%%%%%%%%%%%%%%%%%%%%%%%%%%%%%%%%%%%%%
\section{Massless Spacetime Spectrum}
%%%%%%%%%%%%%%%%%%%%%%%%%%%%%%%%%%%%%%%%%%%%%%%%%%%%%%%%%%%%%%%%%%%%%%%%%%%
In this section, equipped with information of G-flux on the
${\bf T}_5/{\bf Z}_2$ orbifold fixed points,
we determine the complete low-energy
spectra that propagate on $R_{5,1}$. Recall that the M(atrix) theory
describes M theory infinitely boosted along ${\bf R}_{1,1} \in
{\bf R}_{5,1}$.
We follow the same strategy adopted in our previous work~\cite{kimrey} for 
the heterotic M(atrix) theory: 
we determine the low-energy spacetime spectrum by
identifying all possible configurations of probing branes that are
consistent with orbifold projection conditions.

On ${\bf R}_{5,1}$, massless fields are labelled by representations of the
little group  Spin $(4) := SU_L(2) \otimes SU_R (2)$.
States of spin $\le 2$ are:
\bee
{\rm graviton}: &&( {\bf 3,3}) \nonumber \\
{\rm gravitino} : && ( { \bf 3,2}) \,\,\, {\rm or} \, \,\, 
( {\bf 2, 3}) \nonumber \\
{\rm tensor} : && 
({\bf 3,1})_{\rm ASD} \,\,\, {\rm or} \,\,\, ({\bf 1,3})_{\rm SD}
\nonumber \\ 
{\rm vector} : && ({\bf 2,2}) \nonumber \\
{\rm spinor} : && ({\bf 2, 1}) \,\,\, {\rm or} \,\,\, ({\bf 1,2})
\nonumber \\
{\rm scalar} : && ({\bf 1, 1}).
\label{states}
\eee
The chiral $(2,0)$ supergravity in six dimensions, to which we will 
identify the M(atrix) theory compactified on ${\bf T}_5/{\bf Z}_2$ below,
admits two massless super-multiplets. The supergravity multiplet consists
of 
\be
\bf (3, 3) \oplus {\rm 4} (2, 3) \oplus {\rm 5} (1, 3) ,
\ee
while tensor multiplet consists of
\be
\bf (3, 1) \oplus {\rm 4} (2, 1) \oplus {\rm 5} (1,1).
\ee

Below, we will identify all the bosonic field contents using the M(atrix)
theory branes as probes. From the ${\cal N}=8$ supersymmetry, the fermionic
fields content follows automatically and fits into the above multiplet
structure.

We will also explore the moduli space
\be
{\cal M}({\bf T}_5/{\bf Z}_2)
= SO(5, 21; {\bf Z}) \backslash SO(5, 21) / SO(5) \otimes SO(21).
\ee

One of the most interesting feature of M(atrix) theory on 
${\bf T}_5/{\bf Z}_2$ orbifold is that the required sixteen longitudinal
five-branes can be constructed out of zero-brane partons. These longitudinal
five-branes are BPS states of M(atrix) theory~\cite{banksseibergshenker} and
reflect fully dynamical entities of the theory. The fact that we can 
arrange nontrivial {\sl spacetime background} such as orbifold as well as 
excitations on it via composite bound-states of zero-brane partons should 
be viewed as the most interesting and important feature of the M(atrix)
theory.

%%%%%%%%%%%%%%%%%%%%%%%%%%%%%%%%%%%%%%%%%%%%%%%%%%%%%%%%%%%%%%%%%%%%%%%%%%%%
\subsection{Untwisted Sector of (2,0) Supergravity}
%%%%%%%%%%%%%%%%%%%%%%%%%%%%%%%%%%%%%%%%%%%%%%%%%%%%%%%%%%%%%%%%%%%%%%%%%%%%
Untwisted sector can be probed by studying the M(atrix) theory
far away from the fixed points. As in the heterotic M(atrix) theory case, 
excitations of the off-diagonal parts of Eq.(\ref{chanpatoncondition}) are suppressed. Thus 
M(atrix) theory in the untwisted sector is given by gauge group
$G_{\rm untwisted} = U(N) 
\otimes U(N)/{\bf Z}_2$, where the ${\bf Z}_2$ denotes 
orbifold action exchanging
the two gauge groups.

Consider forward scattering in ${\bf R}_{5,1}$ between two zero-brane partons
at impact parameter $|{\bf r}|$ and relative velocity ${\bf v}$.
The scattering amplitude is easily calculated from the M(atrix) Hamiltonian
with gauge group $G_{\rm untwisted}$. For toroidal compactification, the
$2 \rightarrow 2$ scattering amplitude at M(atrix) one-loop
has been calculated by Berenstein
and Corrado~\cite{berensteincorrado}, and we gratefully make use of their 
results. With appropriate change of the normalization and using Euler-MacLaurin
formula
\bee
{\cal A}_{2D0 \rightarrow 2D0}
&=& - 6 \sum_{n_i} \int {d \omega \over 2 \pi} \,{  {\bf v}^4 \over
(\omega^2/T_A + T_A |{\bf r}|^2 + T_A ( n_i R_i )^2 ) }
\nonumber \\
&\approx& - {6 \over T_A} 
 { d^5 y \over {\rm Vol} ({\bf T}_5/{\bf Z})_2 } \! \int {d \omega \over
 2 \pi} \,
{{\bf v}^4 \over ( \omega^2 + T_A^2 |{\bf r}|^2 + y^2) }
\nonumber \\
& = & - { \kappa_{11}^2 \over 2 \pi^3 R_{11}^3} \,
{ 1\over R_5 R_6 R_7 R_8 R_9} \, { {\bf v}^4 \over |{\bf r}|^2}.
\label{longrange}
\eee
Here, $\kappa_{11}$ is the eleven-dimensional gravitational coupling constant
with normalization $L_{11} = R^{(11)} / 2 \kappa_{11}^2 + \cdots$ and
$T_A^3  \equiv  (2 \pi)^4 R_{11}^3 / \kappa_{11}^2$. The result is precisely
the same as the $d=6$ graviton-gravition elastic scattering
amplitude at tree--level and at light--front frame. The scattering amplitude
is ${\bf Z}_2$--even, hence, survives the orbifold projection.
This establishes the existence of $({\bf 3,3})$ graviton propagating in 
${\bf R}_{5,1}$. It is also possible that both of the zero-branes scatter
in forward direction
inside  ${\bf T}_5/{\bf Z}_2$. The scattering amplitude, hence, the
corresponding exchanged graviton state, is again
${\bf Z}_2$--even. The $5 \cdot 6 / 2 = 15$ different graviton
polarizations then give rise to $15 ({\bf 1,1})$ real scalar fields on
${\bf R}_{5,1}$.
Forward scattering between a zero-brane propagating on ${\bf R}_{5,1}$ and
another propagating inside ${\bf T}_5/{\bf Z}_2$ is ${\bf Z}_2$--odd. This
implies that there is no Kaluza-Klein gravi-photon in the spectra.

As an another probe, we use the M(atrix) two-brane made out of 
Landau-orbiting zero-brane partons. The two-brane can orient and propagate
into various directions in ${\bf T}_5/{\bf Z}_2 \otimes {\bf R}_{5,1}$.
Only ${\bf Z}_2$ invariant motions will be left after the orbifold projection.
Consider a two-brane oriented entirely within
${\bf T}_5/{\bf Z}_2$ orbifold. Configuration of this so-called internal 
two-brane is given
by~\cite{kimrey}
\be
A^I = \left(
\begin{array}{cc} Q & 0 \\ 0 & - Q^{\rm T} \end{array} \right)
, \hskip0.5cm
A^J =  \left(
\begin{array}{cc} P & 0 \\ 0 & - P^{\rm T} \end{array} \right)
, \hskip1cm (I, J = 5, \cdots, 9)
\ee
where $P, Q$ are $(N \times N)$ ($N \rightarrow \infty$) Hermitian matrices
satisfying commutation relation $[Q, P] = +1 $. As explained in Section 2, 
the ${\bf Z}_2$ parity is encoded into the BPS central charge density:
\be
{\cal Z}^{IJ} \equiv 
[A^I, A^J] = [Q, P] \oplus [- Q^{\rm T} , - P^{\rm T}]
=   \sigma_3 \otimes {\bf I}_{N \times N}.
\ee
The configuration is ${\bf Z}_2$--odd. To survive the orbifold projetion, 
the internal two-brane has to propagate only inside ${\bf T}_5/{\bf Z}_2$.
Geometrically there are 
$5 \cdot 6\cdot 3 / 3 \cdot 2 = 10$ such distinct propagations. 
Coupled to each of them are 10 $({\bf 1, 1})$
real-valued scalar fields in the spacetime ${\bf R}_{5,1}$.

Next consider a two-brane extended across  both both  ${\bf T}_5/{\bf Z}_2$ and 
${\bf R}_{5,1}$. This is a direct counterpart of the twisted membranes
in the heterotic M(atrix) theory. The configuration is given by~\cite{kimrey}:
\be
A^I = {1 \over \sqrt 2} \left( \begin{array}{cc} Q & 0 \\ 0 & - Q^{\rm T}
\end{array} \right),
\hskip0.5cm
{\bf X}^j = {1 \over \sqrt 2}
\left( \begin{array}{cc} P & 0 \\ 0 & P^{\rm T} \end{array} \right),
\hskip1cm (I = 5, \cdots, 9; \,\,\, j = 1, \cdots, 4).
\ee
The BPS central charge density
\be
{\cal Z}^{IJ} \equiv 
[A^I, {\bf X}^j] = {1 \over 2}  [Q, P] \oplus [- Q^{\rm T}, P^{\rm T}]
= { 1 \over 2} {\bf I}_{2N \times 2N},
\ee
shows that the configuration is ${\bf Z}_2$--even. Therefore, orbifold
projection surviving propagation is within ${\bf R}_{1,1}$, viz., BPS
soliton strings in six dimensions. There are five distinct orientations of
the two-brane inside ${\bf T}_5/{\bf Z}_2$. Since the strings couple to 
antisymmetric tensor fields, we find that the spacetime spectra include
$5({\bf 5,1}) \oplus 5({\bf 1, 3})$ (anti)-self-dual tensor fields.

Yet another possible probe is the longitudinal five-brane\footnote{ 
It is the infinitely boosted D4-brane in strongly coupled type IIA string.}
in the M(atrix) theory.
The configuration can be constructed again out of zero-brane partons in
the M(atrix) theory.
\bee
A^I &=& \left( \begin{array}{cc} Q_1 & 0 \\ 0 & - Q^{\rm T}_1 \end{array}
\right), \hskip0.5cm
A^J =  \left( \begin{array}{cc} P_1 &  0 \\ 0 & - P^{\rm T}_1 \end{array}
\right), \hskip1cm [Q_1, P_1] = + 1
\nonumber \\ 
A^K &=& \left( \begin{array}{cc} Q_2 & 0 \\ 0 & - Q^{\rm T}_2 \end{array}
\right), \hskip0.5cm
A^L =  \left( \begin{array}{cc} P_2 & 0 \\ 0 & - P^{\rm T}_2 \end{array}
\right), \hskip1cm [Q_2, P_2] = + 1.
\label{l5brane}
\eee
The configuration carries the following BPS central charge densities
\bee
{\cal Z}^{IJ} &\equiv& 
[A^I, A^J] = {1 \over 2} \sigma_3 \otimes {\bf I}_{N \times N}
\hskip0.75cm
{\cal Z}^{KL} \equiv 
[A^K, A^L] = {1 \over 2}  \sigma_3 \otimes {\bf I}_{N \times N}
\nonumber \\
{\cal Z}^{IJKL} &\equiv& [A^{[I}, A^J] [A^K, A^{L]} ]
= {\cal Z}^{IJ} {\cal Z}^{KL} =
{1 \over 4} {\bf I}_{2N \times 2N}.
\label{centralcharge} 
\eee
They show that all five distinct wrapping modes on ${\bf T}_5/{\bf Z}_2$
of the longitudinal five-brane are ${\bf Z}_2$--parity even.
Propagation only along ${\bf R}_{5,1}$ 
survives the orbifold projection and gives rise to five BPS soliton string in
spacetime. Again coupled to them are 5 $({\bf 3, 1}) $ and $({\bf 1, 3})$ 
(anti)-self-dual tensor fields\footnote{It is interesting to
note that, while the longitudinal
five-brane constructed as above carries extra two-brane
charges ${\cal Z}^{IJ}, {\cal Z}^{KL}$, each of them are odd under 
${\bf Z}_2$--parity for the above propagation, hence, are 
completely projected out.}.
However, these tensor fields are not independent ones. The longitudinal
five-brane is electric-magnetic dual to the transverse two-brane in
the M(atrix) theory as is evidenced from Berry's phase analysis. 
The BPS strings on ${\bf R}_{5,1}$ reduced from these branes are
electric-magnetic dual. In fact, (anti)-self-dual combinations of
these BPS strings are the ones minimally coupled to the 5 $ ({\bf 3, 1})$
and 5 $({\bf 1, 3})$ tensor fields.   

The above identification of `electric' and `magnetic' BPS strings on
${\bf R}_{5,1}$ point to an enhanced $SU(2)$ symmetry among tensor
fields. A natural pairing of the `electric' and the `magnetic' BPS 
strings is that they are dual on ${\bf T}_5$ lattice. For a fixed
volume of the ${\bf T}_5/{\bf Z}_2$, it is clear that the BPS tension
of the (anti)-self-dual BPS string is given by:
\be
T_{\bf e, m}
= \sum_{i=1}^5 \kappa | e^i R_i \pm {\ell_{11}^2 \over R_i} m^i|.
\ee
Thus, when the radius of any of the five toroidal direction is at
the self-dual point $R_i = \ell_{11}$, the anti-self-dual strings
become tensionless. Clearly, this is the M-theory counterpart of
the momentum-winding duality. At the self-dual point, it then follows that 
there arises an enhanced $SU(2)$ symmetry. 

It is also possible that the longitudinal five-brane wraps less than 
four directions around ${\bf T}_5/{\bf Z}_2$. Inspection of the
central charge density shows that propagation of 
these configurations are either ${\bf Z}_2$--parity odd, hence,
projected out or reduces to the above already analyzed cases.

Combining the $({\bf 3, 3})$ graviton and 5 $ ({\bf 3,1})$ self-dual
tensor fields, we obtain the $(0,2)$ supergravity multiplet. 
Similarly, combining the remaining 5 $({\bf 1, 3})$ anti-self-dual
tensor fields, 15 $  ({\bf 1, 1})$ scalars and 10 $({\bf 1, 1})$
scalars that couple to ${\bf T}_5/{\bf Z}_2$ propagation of the
zero- and two-branes respectively, that couples to zero-brane
propagation inside, we obtain five $(0,2)$ tensor multiplets.

%%%%%%%%%%%%%%%%%%%%%%%%%%%%%%%%%%%%%%%%%%%%%%%%%%%%%%%%%%%%%%%%%%%%%%%%%%%%
\subsection{Turning on Longitudinal Five-Branes}
%%%%%%%%%%%%%%%%%%%%%%%%%%%%%%%%%%%%%%%%%%%%%%%%%%%%%%%%%%%%%%%%%%%%%%%%%%%%
In Section 2, we have shown that each orbifold fixed point carry 
$[G/2\pi] = - 1/2$ G-flux. On ${\bf T}_5/{\bf Z}_2$, the total G-flux then 
adds up to $-16$ units. The anomalous flux has to be cancelled
in order to maintain the $(0,2)$ supersymmetry on ${\bf R}_{5,1}$. 
As we have discussed in Section 2, the cancellation
can be achieved by putting sixteen
longitudinal five-branes on ${\bf R}_{5,1}$. This is
precisely M(atrix) theory counterpart
to the earlier observation of Witten~\cite{wittent5} 
and Dasgupta and Mukhi~\cite{ dasguptamukhi} within the context of 
low-energy eleven-dimensional supergravity approximation.

In M(atrix) theory, ground--state configuration of the zero-brane partons
that corresponds to sixteen longitudinal five-branes is straightforwardly
constructed. The 
configuration is essentially the same as Eq.(37) except that they are
now oriented along ${\bf R}_{5,1}$. Labelling the sixteen five-branes
as paired indices $(a,b)$ where $a,b = 1, \cdots, 16$, 
the zero-brane parton configuration on ${\bf R}_4 \in {\bf R}_{5,1}$
is given by $(32 n + N) \times (32 n + N)$ matrices: 
\be
{\bf X}^i = \left( \begin{array}{ccccc}
{\bf X}_{a=1}^i &&&& \\
& {\bf X}_{a=2}^i &&& \\
&& \ddots && \\
&&& {\bf X}_{a=16}^i & \\
&&&& {\bf Y}^i \\
\end{array} \right).
\label{sixteenbranes}
\ee
Here, ${\bf X}_a$ denotes coordinate matrices for $a-$th longitudinal
five-brane and its orientifold mirror five-brane in terms of $2n \times 2n$ 
matrices:
\bee
{\bf X}^1_a &=& \left( \begin{array}{cc} Q_{1a} & 0 \\ 0 & Q^{\rm T}_{1a}
 \end{array} \right), \hskip0.5cm
{\bf X}^2_b = \left( \begin{array}{cc} P_{1b} & 0 \\ 0 & P^{\rm T}_{1b}
\end{array} \right), \hskip1cm
[Q_{1a}, P_{1b}] = \delta_{ab} \, { 1 \over \sqrt{ 2n}} 
{\bf I}_{2n \times 2n};
\nonumber \\
{\bf X}^3_a &=& \left( \begin{array}{cc} Q_{2a} & 0 \\ 0 & Q^{\rm T}_{2a}
 \end{array} \right), \hskip0.5cm
{\bf X}^4_b = \left( \begin{array}{cc} P_{2b} & 0 \\ 0 & P^{\rm T}_{2b}
\end{array} \right), \hskip1cm
[Q_{2a}, P_{2b}] = \delta_{ab} {1 \over \sqrt{ 2n}} 
{\bf I}_{2n \times 2n}.
\label{eachbrane}
\eee
The coordinates of N unclustered zero-brane partons and their images 
are separately denoted as ${\bf Y}^i$. In the infinite momentum limit
of the M(atrix) theory, both n and N goes to infinity in proportion.

The configuration carries the correct BPS five-brane central charge of +16 
units :
\bee
{\cal Z}^{[12][34]} &\equiv&  \sum_{a=1}^{16} 
{\cal Z}^{[12][34]}_a
[{\bf X}^1, {\bf X}^2]  [{\bf X}^3, {\bf X}^4]
\nonumber \\
&=& 16 \, {1 \over 2 n} {\bf I}_{2n \times 2n}.
\label{l5centralcharge}
\eee
Each of the sixteen configurations labelled $a=1, \cdots, 16$ give
rise precisely to a longitudinal five-brane with a unit charge.

In addition, the configuration carries BPS two-brane central charges:
\bee
{\cal Z}^{[12]}_a &=& [{\bf X}^1, {\bf X}^2]_a =
{1 \over  \sqrt {2n}} {\bf I}_{2n \times 2n}, \hskip1cm (a=1,2,3,4)
\nonumber \\
{\cal Z}^{[34]}_b &=& [{\bf X}^3, {\bf X}^4]_b = 
{1 \over \sqrt {2n}} {\bf I}_{2n \times 2n}, \hskip1cm
(b=1,2,3,4).
\eee
As such, the two-brane BPS central charges diverge as ${\sqrt n}
\rightarrow \infty$.

We now show that turning on longitudinal five-branes as above still
preserves the $(2,0)$ chiral supersymmetry on ${\bf R}_{5,1}$. 
To do so, consider a general longitudinal five-brane configuration
Eqs.(\ref{sixteenbranes},\ref{eachbrane}) 
in which the commutation relations of $P_{a}, Q_{b}$'s are now 
replaced by:
\be
[Q_{1a}, P_{1b}] =  \delta_{ab} A \,\, {\bf I}_{2n \times 
2n}, \hskip0.75cm 
[Q_{2a}, P_{2b}] = \delta_{ab}
B \,\, {\bf I}_{2n \times 2n} , 
\ee
where $ AB = {1 \over 2n}$.
Whether the configuration preserves the $(2,0)$ chiral supersymmetry
is most conveniently probed by scattering off a zero-brane. 
Consider the zero-brane probe moving slowly on ${\bf T}_5
/{\bf Z}_2$:
\be
A^1 = \left( \begin{array}{ccccc} 0 & & & & \\ & 0 & & & \\
  & & 0 & & \\ & & & \ddots & \\ & & & & r \sigma_3  \end{array} \right)
, \hskip0.5cm
 A^2 = \left( \begin{array}{ccccc} 0 & & & & \\ & 0 & & & \\
   & & 0 & & \\ 
   & & & \ddots & \\ & & & & vt \sigma_3 \end{array} \right)
, \hskip0.5cm
A^3 = A^4 = A^5 = 0.
\label{zerobackground}
\ee
For a single longitudinal five-brane (type IIA four-brane), the scattering
has been studied by Lifschytz~\cite{lifschytz}. We generalize his calculation
adopted to the present context. In the background Eqs.(\ref{sixteenbranes}, 
\ref{eachbrane},\ref{zerobackground}), massive states come from the last 
two rows and columns. They are four different stretched 
open strings connecting longitudinal five-branes on one hand
and their images and the probe zero-brane and its image on another end. 
The one-loop
integral gives rise to the phase shift for a zero-brane probe
scattering off sixteen longitudinal five-branes is given by~\cite{lifschytz}:
\bee
\delta_{\rm 16 \,\,\, L5} (r, v) &=& 
- \int_{-\infty}^{+\infty} dt \, {\cal V}_{\rm
16 \,\,\, L5} (r, vt) \nonumber \\
&=&
16  \int {d s \over s} e^{- r^2 s} \, 
( 8 \sinh s A \sinh s B \sin sv)^{-1} \nonumber \\
& \times & [
2 + 2 \cos 2 s v + 2 \cosh 2sA + 2 \cosh 2 s B - 8 \cos sv \cosh sA 
\cosh sB].
\eee
For $A \ne B$, this gives rise to {\sl static} potential:
\be
{\cal V}_{\rm 16 \,\,\, L5} (r, v) \approx
- 16 \cdot {1 \over 16} {1 \over r^3} [ {(A^2 - B^2)^2 \over AB}
+ 2 {(A^2 + B^2) \over AB}  v^2  + \cdots], 
\ee 
hence, 
incompatible with the ${\cal N}=8$ supersymmetry of the orbifold M(atrix)
theory. Consequently, the $(2,0)$ chiral supersymmetry on ${\bf R}_{5,1}$ is
explicitly broken.
On the other hand, if $A = B = 1/ {\sqrt {2n}}$ as in Eq.(\ref{eachbrane}), 
the static potential vanishes identically and the potential
between zero-brane and sixteen longitudinal five-branes is obtained. 
In fact, in this case $A = B = 1/ {\sqrt {2n}}$, the scattering phase shift 
can be written as:
\be
\delta_{\rm 16 \,\,\,\, L5} (r, v) 
= + 16 \int {d s \over s} e^{ - r^2 s}
\Big[ { 1 - \cos s v \over \sin sv} + 
{1 \over 4 \sinh^2 {s /\sqrt{2n}}} 
{3 - 4 \cos sv + \cos 2 sv \over \sin sv} \Big].
\label{susyphaseshift}
\ee
The first term is precisely the contribution of longitudinal five-brane
charge ${\cal Z}^{[12][34]}$, while the second term is the contribution
of ${\cal O}(N)$ zero-branes. Note that the effect
of embedded two-brane charges ${\cal Z}^{[12]}, {\cal Z}^{[34]}$ to 
the phase shift vanishes miraculously in the limit $A = B$, 
corresponding to isotropic longitudinal five-branes.    

The fact that the symmetric configuration $A = B$ is compatible with
the $(2,0)$ chiral supersymmetry on ${\bf R}_{5,1}$ can be also seen from
the supersymmetry transformation rules of the M(atrix) theory.
In the background of the sixteen longitudinal five-branes,  
\be
\delta {\bf \Theta}  
= 16 \cdot {i \over 4} \Big( \Gamma_{12} A + \Gamma_{34} B 
\Big)\, {\bf \epsilon} = 0.
\ee
Therefore, only for the symmetric configuration $A = B$, we find a 
half-supersymmetry-preserving BPS condition
\be
\Gamma_{[12]} = - \Gamma_{[34]} \hskip0.3cm \leftrightarrow \hskip0.3cm
\Gamma_1 \Gamma_2 \Gamma_3 \Gamma_4 = + 1 
\hskip0.3cm \leftrightarrow \hskip0.3cm 
\Gamma_5 \Gamma_6 \cdots \Gamma_9 = + 1.
\ee
The condition is precisely the same spinor projection condition
for the ${\bf T}_5/{\bf Z}_2$ orbifold and for the $(2,0)$ chiral
supersymmetry on ${\bf R}_{5,1}$. We conclude that turning on 
sixteen longitudinal five-branes symmetrically as in Eqs.(\ref{sixteenbranes},\ref{eachbrane}) is compatible with the 
supersymmetry.

We again emphasize the importance of above microscopic construction of
longitudinal five-brane background. This is in complete agreement with
the spirit of the M(atrix) theory: nontrivial spacetime backgrounds 
as well as localized excitations propagating on it are to be built entirely
out of zero-brane partons. That this is a consistent picture should be hardly
surprising. The M-theory graviton is a composite bound-state of zero-brane
partons. Nontrivial spacetime background is nothing but a coherent 
state configuration of the gravitons, hence, in turn, composite bound-state
of zero-brane partons at the microscopic level. Our construction of the
consistent ${\bf T}_5/{\bf Z}_2$ background then should be viewed as 
a realization of nontrivial spacetime out of matrices. 

%%%%%%%%%%%%%%%%%%%%%%%%%%%%%%%%%%%%%%%%%%%%%%%%%%%%%%%%%%%%%%%%%%%%%%%%%%%%
\subsection{Twisted Sector of (2,0) Supergravity}
%%%%%%%%%%%%%%%%%%%%%%%%%%%%%%%%%%%%%%%%%%%%%%%%%%%%%%%%%%%%%%%%%%%%%%%%%%%%
In Section 3, we have already identified the twisted sector
degrees of freedom in the M(atrix) theory from the consideration of
the G-flux conservation and gauge anomaly canellation of ${\cal N} = 1$
supersymmetric gauge theories on a dual torus ${\tilde {\bf T}}_5$.
Having turned on sixteen longitudinal five-branes and cancelled gauge and
supersymmetry anomalies, we now determine the twisted sector spacetime spectrum.

So far we have not specified positions of parallel longitudinal five-branes
on ${\bf T}_5/{\bf Z}_2$. Associated with each longitudinal five-brane are
five coordinates. They transform as $({\bf 5, 1})$ under $SO(5) \otimes
SO(4)$. Therefore, there are 80 real scalar fields  on ${\bf R}_{5,1}$.
Associated also with each longitudinal five-brane is an anti-self-dual
tensor field. Combining them together, we find sixteen tensor multiplets.
Excitations of the sixteen longitudinal five-branes are described by
open two-branes connecting them. Consider the five-branes clustered 
near one of the 32 fixed points. Positions of the sixteen five-branes
on ${\bf T}_5/{\bf Z}_2$ are given by:
\be
A^I = \left( \begin{array}{ccccc}
r_1 \sigma_3 \otimes {\bf I} & & & & \\
& r_2 \sigma_3 \otimes {\bf I} & & & \\
& & \ddots & &  \\
& & & r_{16} \sigma_3 \otimes {\bf I} \\
& & & & 0 \end{array} \right)
\ee
Here, $\sigma_3$ denotes pairing of five-branes with their images across
the fixed point. In this background, fluctuations of the longitudinal
five-branes are described by the off-diagonal parts of the ${\bf X}^i$
$(i = 1,2,3,4)$ matrices. 

What are the possible singularities and enhanced symmetries thereof?
Recall that we have chosen USp(2N) as the gauge group of the
M(atrix) theory on ${\bf T}_5/{\bf Z}_2$~\footnote{
Note that background configurations
of transverse two-branes and longitudinal five-branes are independent of
the choice of the gauge groups. It is only when we examine enhanced
symmetries that distinguishes the SO(2N) and USp(2N). }
Inferring from the canonical matrix structure, we find that there arises two
possible enhanced symmetries in the twisted sector, which turns out
to be
the direct counterpart of $A_{m-1}$ and $D_m$ singularities $m \le 16$
of type IIB string compactified on K3.

The $A_{m-1}$ singularities arise when
the five-branes coalesce on a point on ${\bf T}_5/{\bf Z}_2$ away from
the fixed point. In this case, the off-diagonal elements in the diagonal
sub-matrices become massless. Since the diagonal sub-matrices are 
unconstrained Hermitian matrices, we find that the enhanced symmetry is
$(U(m) \times U(m)) / {\bf Z}_2$, where each of the two $U(m)$'s are
associated with overlapping five-branes and their orientifold images,
and the ${\bf Z}_2$ projects out into ${\bf Z}_2$--parity even 
configurations of the two clusters. The overlap configuration is achieved by 
tuning $5m - 5$ parameters. Hence, the loci of $A_{m-1}$ type singularity
and associated $U(m)$ enhanced symmetry is a hypersurface of co-dimension
$5 (17 - m)$.

Similarly the $D_m$ singularities arise when the five-branes coalesce on
one of the fixed points. In this case, in addition to the off-diagonal
elements in the block-diagonal Hermitian matrices, the off-diagonal
Hermitian, antisymmetric matrices become massless and excited. Since all
of the off-diagonal entries of ${\bf X}^i$ are excited, the resulting
enhanced symmetry is given by the group in which ${\bf X}^i$ are in the
adjoint representations. Recall that we have chosen the M(atrix) theory
gauge group to be USp(2N). In this case, the ${\bf X}^i$'s are in
antisymmetric representations of USp(2N). Hence, we identify that 
the symmetry group of which the ${\bf X}^i$'s are in adjoint 
representations is SO(2m).
We find that the enhanced symmetry when $m$ $(m \le 16)$
longitudinal five-branes approach one of the fixed points there arises
an enhanced global symmetry group of $D_m$ type, SO(2m). This is
precisely what is known to take place for type IIB string compactified
on K3 with $D_m$ type singularities.

%%%%%%%%%%%%%%%%%%%%%%%%%%%%%%%%%%%%%%%%%%%%%%%%%%%%%%%%%%%%%%%%%%%%%%%%%%%%
\section{Further Toroidal Compactifications}
%%%%%%%%%%%%%%%%%%%%%%%%%%%%%%%%%%%%%%%%%%%%%%%%%%%%%%%%%%%%%%%%%%%%%%%%%%%%
It is of interest also to study higher-dimensional compactification in
which ${\bf T}_5/{\bf Z}_2$ forms a subspace. The simplest example is
a direct product compactification $({\bf T}_5/{\bf Z}_2) \otimes {\bf T}_n$
for $n = 1, 2, 3, 4$. They all possess ${\cal N} = 8$ supersymmetries.
What would be corresponding M(atrix) theory description to these
compactifications?
%%%%%%%%%%%%%%%%%%%%%%%%%%%%%%%%%%%%%%%%%%%%%%%%%%%%%%%%%%%%%%%%%%%%%%%%%%%%
\subsection{M(atrix) Gauge Theory on Dual Orbifold ${\tilde {\bf T}}_5
\otimes ({\tilde {\bf T}}_n/{\bf Z}_2)$}
%%%%%%%%%%%%%%%%%%%%%%%%%%%%%%%%%%%%%%%%%%%%%%%%%%%%%%%%%%%%%%%%%%%%%%%%%%%%
The starting point is the M(atrix) theory in the covering space
${\bf T}_{5+n}$ of radii $R_1, \cdots, R_{5+n}$ 
and arrange a mirror parton to every zero-brane parton. 
In the small volume limit, the parton dynamics in the covering space
is governed by ${\cal N} = 16$ supersymmetric U(2N) Yang-Mills theory. 
The parameter space ${\tilde {\bf T}}_{5+n}$ on which the Yang-Mills theory 
is defined is dual to the covering space ${\bf T}_{5+n}$. 
Denote their radii as ${\tilde R}_1, \cdots, {\tilde R}_{5+n}$. 
These radii and Yang-Mills gauge 
coupling constant are related to the M-theory parameters $R_1, \cdots,
R_{5+n}, R_{11}$ and $\ell_{11}$ as:
\be
g^2_{\rm YM} = \ell_{11}^3 \,\, \Big( 
{\rm Vol^{n+2}({\tilde {\bf T}}_{5+n}) \over 
{\rm Vol}^3({\bf T}_{5+n}) }
\Big)^{1 \over 5+n} ,
\hskip1.5cm
{\tilde R}_M = {\ell_{11}^3 \over R_M R_{11}}.
\ee
Denote the 
(5+n+1) dimensional fields as $A_M (x)$ ($M = 0, 1, \cdots, 5+n$) 
for the gauge field, ${\bf X}^i$ ($i=5+n+1, \cdots, 9$) for the pseudo-scalar 
adjoint scalar field and ${\bf \Theta}_\alpha$ for the 16-component spinor. 
We now construct M(atrix) gauge theory relevant to
$({\bf T}_5/{\bf Z}_2 ) \otimes 
{\bf T}_n$ orbifold by identifying a suitable involution and projection 
thereof. Acting on the above fields, define a `parity' transformation $P$ 
defined by
\bee
P \hskip0.5cm : \hskip0.5cm
A_M (x) &\rightarrow& P \cdot A_M (x)
\cdot P^{-1} \,\,\,
\equiv  {P_M}^N \, A_N (P \cdot x) \nonumber \\
{\bf X}^i (x) &\rightarrow& P \cdot {\bf X}^i(x) \cdot P^{-1}
\,\,\, \equiv \,\, - \,\,  {\bf X}^i(P \cdot x)
\nonumber \\
{\bf \Theta}_\alpha (x) &\rightarrow&  P \cdot {\bf \Theta}_\alpha (x)
\cdot P^{-1} \,\,\, \equiv \,\,\, \Gamma_\perp {\bf \Theta} (P \cdot x).
\eee
Here, the `parity' matrix acting on (5+n+1)-dimensional parameter space
coordinates $x^\mu$'s is defined by 
\be
{P_\mu}^\nu = {\rm diag.} (+, + , \cdots, + , -, \cdots, -), 
\ee
where the negative signs are for 6-th through (5+n)-th
entries. The transformation coincides with the conventional definition of 
parity for (5+n+1) = odd but not for even. We will nevertheless call the
transformation as `parity' transformation for lack of better notation.
Note that we have taken ${\bf X}^i$'s as pseudo-scalar fields,
as is always the case for those originating from dimensional reduction. 
In addition, define an orientation reversal transormation 
$\Omega$, whose local action on the fields is given by:
\bee
\Omega \hskip0.5cm : \hskip0.5cm
A_\mu (x) &\rightarrow& \Omega \cdot A_\mu(x) \cdot \Omega^{-1}
\,\,\, \equiv \,\,\, \pm \,\, A_\mu^{\rm T}(x)
\nonumber \\
{\bf X}^i(x) &\rightarrow& \Omega \cdot {\bf X}^i (x) \cdot \Omega^{-1}
\,\,\, \equiv \,\,\, \pm \,\, {\bf X}^{i \rm T} (x) \nonumber \\
{\bf \Theta}_\alpha (x) &\rightarrow& \Omega \cdot {\bf \Theta}_\alpha^{\rm T}
(x) \cdot \Omega^{-1} \, \equiv \, \pm \,\, \epsilon {\bf \Theta}_\alpha^{\rm T}(x).
\eee
Here, the overall and the Bose-Fermi relative sign ambiguity is as before.
The two-fold ambiguity ($\pm$) 
gives rise to USp(2N) and SO(2N) M(atrix) gauge group
for $\pm $ sign choices respectively. Extending the argument in sections
2 and 3, we choose the M(atrix) gauge group to be USp(2N). 
The relative two-fold ambiguity ($\epsilon = \pm$)
between the bosonic and the fermionic fields amounts to the interchange
between the two eigenstates of $\Gamma_\perp$, hence, is fixed by the
gamma matrix representations. 

It is straightforward to see that combined action $\Pi \equiv \Omega \cdot
P$ is an invariance of the covering space M(atrix) gauge theory. 
Therefore, projecting out with $(1 + \Pi)/2$, we obtain M(atrix) gauge theory 
for $({\bf T}_5/{\bf Z}_2) \otimes {\bf T}_n$ orbifold compactification of
the M theory. The resulting orbifold M(atrix) gauge theory has many 
exotic features not encountered in M(atrix) gauge theory obtained after
toroidal compactifications.
First, from the action of $P$, we find that the parameter space on which the
M(atrix) gauge theory is defined is a dual orbifold ${\tilde {\bf T}}_5
\otimes ({\tilde {\bf T}}_n/{\bf Z}_2)$. The dual orbifold has $2^n$
fixed boundaries. Each boundary is given by five-dimensional dual torus
${\tilde {\bf T}}_5$. In fact, (5+1)-dimensional chiral gauge theory on the 
boundary can be viewed as the intersection between $(5+p)$-brane and 
$(9-p)$ brane. 

%%%%%%%%%%%%%%%%%%%%%%%%%%%%%%%%%%%%%%%%%%%%%%%%%%%%%%%%%%%%%%%%%%%%%%%%%%%%%
\subsection{Wilson Lines and Gauge Anomaly Cancellation}
%%%%%%%%%%%%%%%%%%%%%%%%%%%%%%%%%%%%%%%%%%%%%%%%%%%%%%%%%%%%%%%%%%%%%%%%%%%%%
We have seen that the orbifold M(atrix) gauge theory is defined on dual
orbifold parameter space. Localized at the $2^n$ five-dimensional fixed 
boundaries are (5+1)-dimensional USp(2N) chiral gauge theories. 
The thirty-two `half' fundamental representation hypermultiplets identified 
as the twisted sector in the previous sections are then redistributed among
the $2^n$ fixed boundaries. In this subsection, we show that transverse
positions of the thirty-two `half' fundamental representation hypermultiplets 
on the parameter space orbifold 
${\tilde {\bf T}}_5 \otimes ({\tilde {\bf T}}_n /{\bf Z}_2 )$ 
is the M(atrix) gauge theory realization of turning on Wilson lines for the
symmetry group associated with the sixteen longitudinal five-branes.

The situation is again exactly the same as the M(atrix) theory on 
$({\bf S}_1/{\bf Z}_2)\otimes{\bf T}_m$ orbifold~\cite{kabatrey}. 
Therefore, for comparison, we briefly recapitulate this case first.  
It has been shown that
the M theory compactification dynamics is described by so-called
heterotic M(atrix) gauge theory on dual orbifold parameter space 
${\tilde {\bf S}}_1 \otimes ({\tilde {\bf T}}_m/{\bf Z}_2)$. Localized at the
$2^m$ fixed boundaries are (1+1)-dimensional SO(2N) chiral gauge theories.
The thirty-two Majorana-Weyl fermionic supermultiplets identified as the
twisted sector are then redistributed among the $2^m$ fixed boundaries.
Furthermore, it has been shown that locations of 
fermionic supermultiplets can be deformed away from the orbifold 
boundaries on the dual orbifold parameter space while maintaining gauge 
anomaly cancellation. A crucial ingredient for the deformation to be possible
is the topological Wess-Zumino coupling of the M(atrix) gauge theory to the
background Ramond-Ramond fields and accompanying modification of
background NS-NS fields. 
By detailed computation, it has been shown that the location deformation is 
nothing but the M(atrix) gauge theory realization of turning on Wilson lines 
for $E_8 \times E_8$ gauge group of toroidally compactified heterotic string. 

We thus begin with the role of topological Wess-Zumino coupling with 
background Ramond-Ramond fields for gauge anomaly cancellation via 
anomaly inflows. For N coincident D-branes, the Wess-Zumino coupling 
is given by
\be
L_{\rm WZ} = \sum_{\rm RR} \int C_{\rm RR} \wedge {\rm Tr} 
\exp \big({i F \over 2 \pi} \big) \sqrt{ \hat A(R)}
\label{wzcoupling}
\ee
with
\be
\sqrt{\hat A(R) } = 1 + {1 \over 2} \hat A_4 + ({1 \over 2} \hat A_8
- {1 \over 8} \hat A^2_4) + \cdots.
\ee
Recall that the orbifold fixed boundaries carry nonzero Ramond-Ramond
charges. The topological Wess-Zumino coupling Eq.(\ref{wzcoupling})
then tells us that the effective action of the orbifold M(atrix) gauge 
theory is anomalous
\be
D_\mu {\delta \Gamma_{\rm M} [A] \over \delta A_\mu}
= {\mu_4 \over 2^m}  \sum_{a=1}^{2^m} 
\int_{\tilde {\bf T}_5 \otimes {\bf R} }
{\rm Tr} (F \wedge F \wedge F). 
\ee 
When the positions of the ${\tilde {\bf T}}_5$ hypersurfaces on which
the `half' fundamental representation hypermultiplets live are symmetrically
located right at $2^n$ ($n=1,\cdots,4$) dual orbifold boundaries,   
they induce gauge anomaly in such a way they cancel the gauge anomaly
present at the orbifold boundaries. In other words, the 
Ramond-Ramond gauge fields are cancelled globally and there is no Wess-Zumino 
coupling present in the corresponding M(atrix) gauge theory. 
Suppose we now move the twisted sector hypermultiplet locations are deformed 
away from the dual orbifold boundaries along ${\tilde {\bf T}}_m/{\bf Z}_2$
directions. In terms of underlying M-theory configurations, this deformation
corresponds to turning on Wilson lines of the M(atrix) gauge fields around the 
compactified ${\bf T}_m$ toroidal directions.   

The relevant Wess-Zumino couplings are 
\bee
L_{\rm WZ} = \int \Big[ \!\!\! && \!\!\!
C_2 \wedge {\rm Tr} (F \wedge F \wedge F \wedge F)
\nonumber \\
&+& C_6 \wedge {\rm Tr} (F \wedge F) \,\, \Big] . 
\eee
Using the equation of motion for the Ramond-Ramond gauge fields, we 
find that these coupling gives rise to 
\be
D_\mu {\delta \Gamma \over \delta A_\mu} 
= \sum {\rm Tr}_{\tt R} (Q \, F \wedge F \wedge F)
\ee
under the gauge transformation.
For the fundamental representation, we find that this is precisely the
needed couplings. 

With the gauge anomalies cancelled locally, it is then possible to 
deform the position of the twisted sector fundamental representation
hypermultiplets. They are indeed in one to one correspondence with the
Wilson lines that break the symmetry group associated with the
sixteen longitudinal five-branes.

%%%%%%%%%%%%%%%%%%%%%%%%%%%%%%%%%%%%%%%%%%%%%%%%%%%%%%%%%%%%%%%%%%%%%%%%%%%%
\section{Discussion}
%%%%%%%%%%%%%%%%%%%%%%%%%%%%%%%%%%%%%%%%%%%%%%%%%%%%%%%%%%%%%%%%%%%%%%%%%%%%
In this paper, continuing the previous investigation, we have studied
in detail M(atrix) theory compactification on ${\bf T}_5/{\bf Z}_2$ orbifold. 
We have shown that the compactification is described by (0,8) supersymmetric
USp(2N) chiral gauge theory coupled to one untwisted sector hypermultiplet
in anti-symmetric representation and thirty-two twisted sector hypermultiplets
in `half' fundamental representations. The gauge theory is defined on the
dual parameter space ${\tilde {\bf T}}_5$. As the volume of the orbifold
gets large, the dual parameter space becomes small. The M(atrix) gauge
theory then reduces to ${\cal N} = 8$ M(atrix) quantum mechanics. 

Similar to ${\bf S}_1/{\bf Z}_2$ orbifold case, the thirty-two fixed
points of ${\bf T}_5/{\bf Z}_2$ turn out to carry anomalous G-flux.
We have verified this from various consistency conditions. First
 parton scattering
off orbifold fixed point has shown unambiguously the presence of 
-1/2 units of G-flux from each fixed point. Second, for shrinking orbifold,
we have found that the (5+1)-dimensional USp(2N) M(atrix) gauge theory
is free from both perturbative and global gauge anomaly
only if we introduce thirty-two `half' hypermultiplets. By pairing
them we have found that the twisted sector hypermultiplets represent the 
sixteen longitudinal five-branes turned on ${\bf R}_{5,1}$.

One of the most interesting feature of M(atrix) theory on ${\bf T}_5/{\bf Z}_2$
orbifold is that the required sixteen longitudinal five-branes can be arranged
out of Landau-orbiting zero-branes. In other words, while introduced
by hand the twisted sector at the start to meet the consistency conditions, 
the required sixteen five-branes can be built out of zero-brane partons.
This is in complete agreement with the idea of M(atrix) theory. Not only
dynamical entities such as strings but also consistent compactification 
background can be built up out of zero-brane partons. 
This is the most distinguishing feature of ${\bf T}_5/{\bf Z}_2$
compared to other compactifications such as ${\bf S}_1/{\bf Z}_2$. 
The required longitudinal five-branes in the former are well-defined BPS 
states of the M(atrix) theory, while the required longitudinal nine-branes
in the latter are not so, as has been shown in \cite{banksseibergshenker}.

Furthermore, we 
have identified the ${\bf R}_{5,1}$ spacetime spectrum entirely
within M(atrix) theory approach. We have found that the resulting massless
excitations are in complete agreement with the six-dimensional (2,0) 
supergravity field content.

We have also shown, upon further toroidal compactification to 
$({\bf T}_5/{\bf Z}_2 ) \otimes {\bf T}_m$ $(m=1,2,3,4)$ orbifolds,
M theory parton dynamics is described by orbifold M(atrix) gauge theory
defined on a dual orbifold ${\tilde {\bf T}}_5 \otimes ({\tilde {\bf T}}_m
/{\bf Z}_2)$. The toroidal compactification direction allows to turn on
Wilson lines and trigger symmetry breaking for the twisted sector gauge
group.

\vskip0.5cm
We are grateful to  K. Intriligator, D. Kabat and E. Witten for 
invaluable discussions.

%%%%%%%%%%%%%%%%%%%%%%%%%%%%%%%%%%%%%%%%%%%%%%%%%%%%%%%%%%%%%%%%%%%

\end{document}